\documentclass[preprint,12pt]{elsarticle}
\usepackage{url} 
\usepackage{amssymb}
\usepackage{amsmath}
\usepackage{float}
\usepackage{booktabs}
\usepackage{array}
\usepackage{tabularx}
\usepackage{multirow}
\usepackage{xcolor}
\usepackage{soul}

\journal{Journal of Power Sources. Accepted Version.}

\begin{document}

\begin{frontmatter}

\title{Optimizing Parameter Estimation for Electrochemical Battery Model: A Comparative Analysis of Operating Profiles on Computational Efficiency and Accuracy}


\author[inst1,inst2,inst3]{Feng Guo \corref{cor1}}
\author[inst1,inst2]{Luis D. Couto}
\author[inst1,inst2]{Khiem Trad}
\author[inst1,inst2]{Grietus Mulder}
\author[inst1,inst2,inst4]{Keivan Haghverdi}
\author[inst1,inst2]{Guillaume Thenaisie}

\affiliation[inst1]{organization={VITO},
            addressline={Boeretang 200}, 
            city={Mol},
            postcode={2400}, 
            country={Belgium}}

\affiliation[inst2]{organization={EnergyVille},
            addressline={Thor Park 8310}, 
            city={Genk},
            postcode={3600}, 
            country={Belgium}}

\affiliation[inst3]{%
    organization={Institute for Materials Research (IMO-imomec), Hasselt University},%
    addressline={Martelarenlaan 42},%
    city={Hasselt},%
    postcode={3500},%
    country={Belgium}%
}
            
\affiliation[inst4]{organization={Institute of Physical Chemistry, RWTH Aachen University},
            city={ Aachen},
            postcode={2074}, 
            country={Germany}}

\cortext[cor1]{Corresponding author:feng.guo@vito.be(Feng Guo). This manuscript has been peer-reviewed and accepted for publication in the Journal of Power Sources. This arXiv version is the accepted manuscript. The final published version is available at the official journal site: https://doi.org/10.1016/j.jpowsour.2025.239044. Please cite the published version. }

\begin{abstract}
Parameter estimation in electrochemical models remains a significant challenge in their application. This study investigates the impact of different operating profiles on electrochemical model parameter estimation to identify the optimal conditions. In particular, the present study is focused on Nickel Manganese Cobalt Oxide(NMC) lithium-ion batteries. Based on five fundamental current profiles (C/5, C/2, 1C, Pulse, DST), 31 combinations of conditions were generated and used for parameter estimation and validation, resulting in 961 evaluation outcomes. The Particle Swarm Optimization is employed for parameter identification in electrochemical models, specifically using the Single Particle Model (SPM). The analysis considered three dimensions: model voltage output error, parameter estimation error, and time cost. Results show that using all five profiles (C/5, C/2, 1C, Pulse, DST) minimizes voltage output error, while {C/5, C/2, Pulse, DST} minimizes parameter estimation error. The shortest  time cost is achieved with {1C}. When considering both model voltage output and parameter errors, {C/5, C/2, 1C, DST} is optimal. For minimizing model voltage output error and time cost, {C/2, 1C} is best, while {1C} is ideal for parameter error and time cost. The comprehensive optimal condition is {C/5, C/2, 1C, DST}. These findings provide guidance for selecting current conditions tailored to specific needs.
\end{abstract}


\end{frontmatter}

\section{Introduction}

In recent years, lithium-ion batteries have emerged as a cornerstone technology, demonstrating unprecedented versatility across diverse applications, from portable electronics to electric vehicles and large-scale energy storage systems\cite{Nyamathulla2024}. This widespread adoption has catalyzed an exponential growth in research focusing on battery modeling, state estimation, and control strategies \cite{Guo2024}. Among these research directions, precise battery modeling has become particularly crucial, as high-fidelity models serve as fundamental tools for accurate state estimation \cite{Wang2023,Zhou2023}, degradation prediction \cite{Mehta2024}, thermal runaway prevention\cite{Alkhedher2024}, and targeted battery management strategies\cite{Guo2024}. While various modeling approaches exist, electrochemical models based on fundamental reaction mechanisms have garnered significant attention from both academic researchers and industry practitioners. These models offer superior accuracy and provide valuable insights into otherwise unmeasurable internal physical parameters and states of the battery\cite{Guo2024}. 

Battery modeling methodologies can be categorized into three main approaches: equivalent circuit models (ECMs), electrochemical models, and machine learning models, each offering distinct advantages for different applications \cite{Zhou2021}. ECMs represent battery behavior through an arrangement of electrical components such as voltage sources, resistors, and capacitors, providing a simplified abstraction of the battery's electrical characteristics \cite{Guo2020}. These models have gained widespread adoption due to their computational efficiency and straightforward implementation in state estimation and control system design \cite{Plett2004,Guo2019,Chen2020}. However, ECMs' fundamental limitation lies in their empirical nature; they lack the capability to capture internal physicochemical processes, making it challenging to accurately predict phenomena such as aging mechanisms and thermal runaway \cite{Guo2024}.

More recently, machine learning models have emerged as a promising alternative, offering a data-driven approach to battery modeling \cite{Heinrich2021}. These models, including neural networks, support vector machines, Gaussian process regression, among others, can capture complex nonlinear relationships in battery behavior without requiring detailed knowledge of underlying physicochemical processes \cite{Zhou2021}. Machine learning models have shown particular success in predicting battery State of Charge(SOC), remaining useful life, and capacity degradation patterns \cite{Raoofi2023,Zhao2023}. However, their accuracy heavily depends on the quality and quantity of training data, and they may struggle to extrapolate beyond their training range.

Electrochemical models, particularly the Pseudo-Two-Dimensional (P2D) model \cite{Fuller1994} and its variants, offer a more rigorous approach based on fundamental electrochemical principles. These models excel in providing detailed insights into internal battery states and demonstrate superior prediction accuracy \cite{Guo2024}. The P2D model, however, requires solving complex systems of partial differential equations, resulting in significant computational overhead. To address this limitation, researchers developed the Single Particle Model (SPM), which simplifies the P2D framework by representing each electrode as a single representative particle and neglecting electrolyte dynamics \cite{Haran1998}. While this simplification substantially reduces computational complexity, it compromises accuracy, particularly under high-current operations where electrolyte effects become significant. This led to the development of the Single Particle Model with Electrolyte Dynamics (SPMe), which maintains some computational advantages while incorporating crucial electrolyte effects \cite{Marquis2019}. Nevertheless, the SPM has been widely adopted owing to its computational efficiency and structural simplicity.

However, the implementation of electrochemical models presents significant challenges, particularly in terms of computational complexity and parameter estimation. The parameter identification process is especially demanding because many crucial parameters cannot be directly measured through experimental methods, necessitating the use of sophisticated optimization algorithms for their estimation. The high-dimensional parameter space of electrochemical models, coupled with the interdependence of these parameters, results in computationally intensive estimation processes that can require substantial time and computational resources. Moreover, electrochemical models typically exhibit overparameterization, a phenomenon where distinct parameter sets can produce identical model outputs, leading to challenges in parameter identifiability and uniqueness of solutions \cite{Miguel2021,andersson2022parametrization}.

In addressing these challenges, several parameter estimation methodologies have been widely adopted in the field. Particle Swarm Optimization (PSO) \cite{Zheng2016,Fan2020}, Least Squares methods \cite{Couto2024}, and Genetic Algorithms (GA) \cite{Feng2020,Sun2022} represent the predominant approaches for parameter identification in electrochemical battery models. Our previous research conducted a comprehensive comparative analysis of these three methodologies in the context of electrochemical model parameter identification \cite{Guo2024b}. The findings demonstrated that while PSO exhibited high accuracy and robustness compared to other methods. Beyond the three mainstream approaches, a variety of alternative metaheuristic algorithms have also been applied to electrochemical model parameter estimation. For instance, Tian et al. introduced the Two-Population Grey Wolf Optimization (TPGWO) for P2D model identification \cite{tian2025physics}. Huang et al. enhanced the standard PSO with a Cuckoo Search strategy, proposing a multi-step Cuckoo Particle Swarm Optimization (MCPSO) to accelerate convergence in SPMe parameterization \cite{huang2024novel}. Tim at al. proposed a strategically switched metaheuristic (SSM) framework, where a convolutional neural network dynamically recommends the most suitable metaheuristic and resampling is used to avoid local optima \cite{kim2023strategically}. Nevertheless, PSO is commonly adopted as a benchmark method for comparison in such studies.

Parameter identification strategies can further be categorized into single-shot identification and stepwise identification \cite{Rojas2024}. In single-shot identification, all parameters are optimized simultaneously using the full dataset and a global objective function, ensuring consistency and reducing the complexity of variable dependencies. In contrast, stepwise identification decomposes the problem into multiple sequential stages, often starting with parameters that are more easily observable (e.g., ohmic resistance or capacity) and subsequently refining the more complex kinetic and diffusion-related parameters \cite{couto2025physics}. While stepwise identification can sometimes improve convergence and interpretability, it typically requires multiple runs of the estimation algorithm, parameter sensitivity analysis, and careful control of operating conditions at each stage, which substantially increases computational cost and experimental complexity.

The selection of operating profiles for parameter estimation plays a crucial role in the accuracy and efficiency of battery parameter identification. Existing research has primarily relied on standardized battery characterization tests, including Constant-Current-Constant-Voltage (CCCV) charging/discharging at various C-rates, pulse tests, and dynamic stress tests (DST). A comprehensive review of the literature reveals diverse approaches to profile selection for parameter estimation \cite{Rojas2024}. Several researchers have explored the testing protocols. Jokar et al. employed Constant-Current discharge data at multiple C-rates (C/10, C/2, 1C, 2C, 5C) for parameter estimation \cite{Jokar2016}. Speltino et al. utilized a combination of constant-current (CC) discharge, charge, and pulse tests \cite{Speltino2009}. Zhang et al. integrated CCCV tests with DST profiles \cite{Zhang2014}. Fan developed a novel hybrid approach combining CC discharge (SOC 100-50\%) with dynamic profiles (SOC 50-0\%), though this non-standardized protocol requires specific experimental setup \cite{Fan2020}. Zhang et al. combined Electrochemical Impedance Spectroscopy (EIS) data with CCCV profiles, although EIS testing requires additional specialized equipment \cite{Zhang2013}. Namor et al. proposed a parameter grouping strategy based on physical significance, utilizing different test profiles for each group. For instance, low-rate CC discharge data were used to estimate capacity-related parameters \cite{Namor2017}. Some researchers also have incorporated parameter sensitivity analysis into the identification process. Park et al. categorized battery parameters into five groups based on sensitivity analysis, subsequently employing different operating profiles for each parameter group \cite{Park2018a,Park2018b}. However, these grouped estimation approaches often require multiple iterations of parameter estimation algorithms, resulting in extended computation times. Chun and Han employed a reinforcement learning approach to design specific current profiles for battery parameter estimation \cite{chun2024maximizing}; however, this method requires separate experimental tests under specially designed operating conditions.

Theoretically, employing a greater variety of testing profiles can improve the accuracy of parameter estimation. However, such an approach inevitably increases both the computational burden and the experimental effort required for data acquisition, thereby extending the overall estimation process. Existing studies have not systematically addressed the trade-off between accuracy and efficiency, leaving a research gap in determining which combinations of commonly used testing profiles can achieve reliable accuracy while minimizing computational cost. 

To bridge this gap, the present study provides one of the first systematic evaluations of testing profile combinations for electrochemical model parameter estimation. Specifically, we investigate five widely adopted battery testing protocols—CCCV charging/discharging at different rates (C/5, C/2, 1C) and dynamic profiles (pulse charging/discharging, DST). By generating 31 distinct combinations of these protocols, we conduct a comprehensive comparative analysis to quantify both estimation accuracy and computational efficiency. It is worth emphasizing that all testing profiles employed in this study are standard laboratory procedures routinely used for model parameterization and validation in both research and industrial battery testing. These protocols effectively capture the essential behaviors observed under practical charging, discharging, and dynamic operating conditions, ensuring that the findings are highly relevant and directly applicable to real-world parameter estimation workflows.
This systematic approach enables the identification of optimal profile combinations that ensure robust accuracy with reduced estimation time, thereby providing practical guidance for accelerating electrochemical model development and deployment in battery management applications.

The structure of this paper is as follows. Section 2 introduces the methodology, including the SPM and the experimental design covering 31 testing profile combinations. Section 3 presents and discusses the results, highlighting the trade-offs between estimation accuracy and computational efficiency, and identifying optimal profile combinations for different applications. Section 4 concludes the study by summarizing the key contributions and practical implications for battery state estimation.

\section{Method}
\begin{figure}[H]
    \centering
    \includegraphics[width=\textwidth]{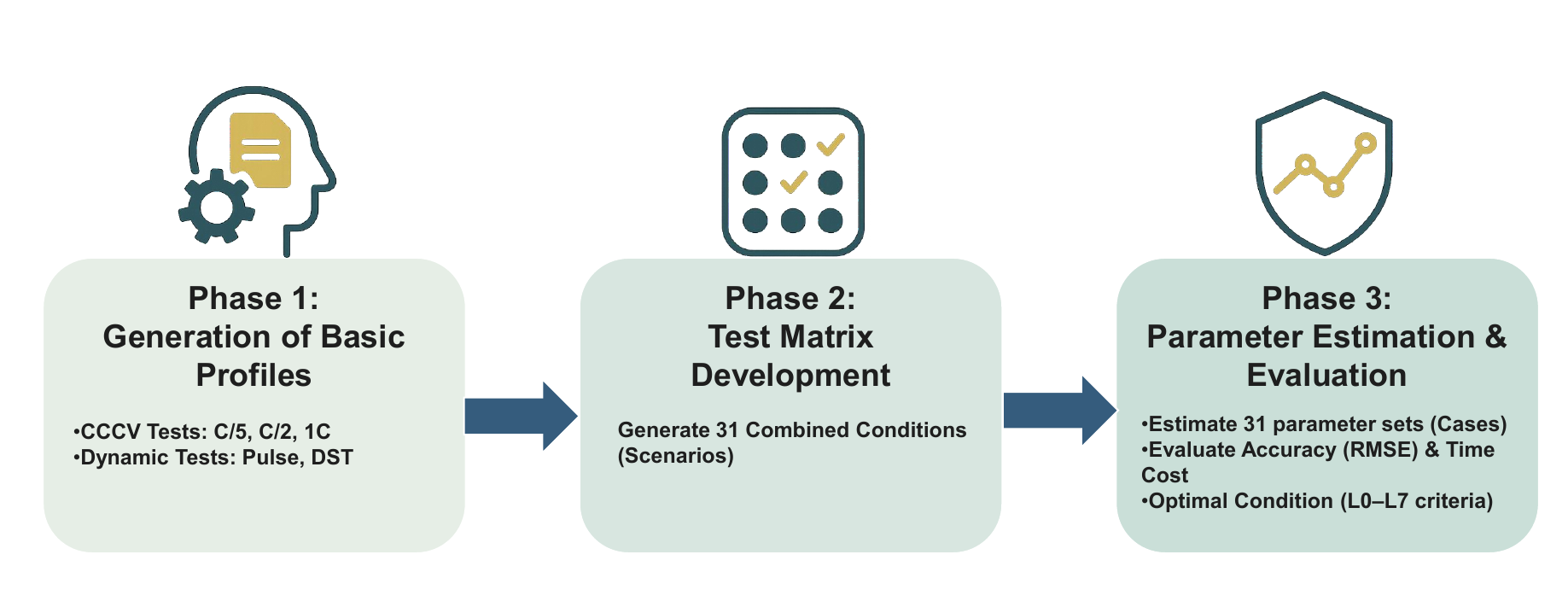}
    \caption{Schematic diagram of the research framework.\label{figure1}}
\end{figure}

This section presents a comprehensive methodology for optimizing battery model parameter estimation through a systematic three-phase approach. As illustrated in Figure~\ref{figure1}, the experimental framework begins with Phase 1, where fundamental test profiles are generated using an SPM battery model. These profiles encompass both CCCV dishcharging and charging at various rates (C/5, C/2, and 1C) and dynamic stress patterns (Pulse and DST), providing a foundation for subsequent analysis. In Phase 2, these basic profiles are systematically combined to create a comprehensive test matrix comprising 31 distinct testing conditions, enabling thorough exploration of various operational Scenarios. Phase 3 focuses on parameter estimation and evaluation, where each test condition is assessed based on two critical metrics: estimation accuracy and computational efficiency. This structured approach ensures a rigorous investigation of the relationship between test conditions and parameter estimation performance, ultimately leading to the identification of optimal testing protocols for battery model parameterization. The following sections provide detailed descriptions of each phase.

\subsection{Generation of Basic Profiles}
Given its computational efficiency and simpler structure relative to the P2D model, the SPM has been extensively studied and applied in engineering practice \cite{Guo2024}; thus, it is selected in this study.

\subsubsection{SPM}
\begin{figure}[H]
\includegraphics[width=\textwidth]{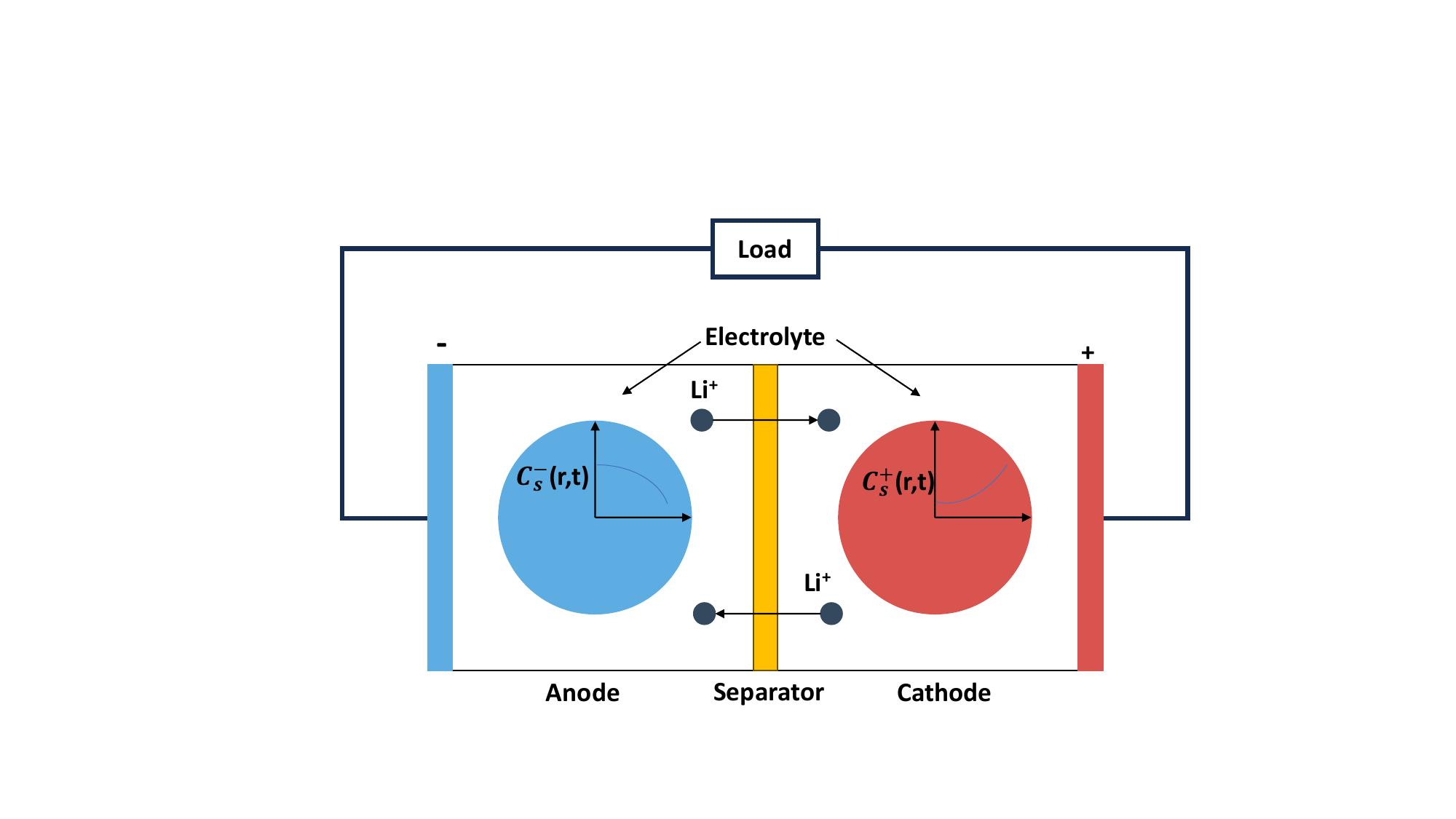}
\caption{Schematic diagram of the SPM.\label{figure2}}
\end{figure}

This study is based on an SPM (see Figure~\ref{figure2}). The model used in this study is derived from our previous research, and further details can be found in Reference \cite{guo2025control,guo5344067cpg}. The model describes the lithium-ion concentration dynamics in battery electrodes through a system of partial differential equations.

\unskip

 The fundamental equation governing lithium concentration in the solid phase for both electrodes is:
\begin{equation}
\frac{\partial c_s^{\pm}}{\partial t}(r,t) = \frac{D_s^{\pm}}{(R_s^{\pm})^2} \frac{\partial}{\partial r} \left( r^2 \frac{\partial c_s^{\pm}}{\partial r}(r,t) \right),\label{eq:1}
\end{equation}
where the superscript $\pm$ denotes positive (+) and negative (-) electrodes respectively. Here, $c_s^{\pm}$ represents the solid-phase lithium concentration, $D_s^{\pm}$ is the diffusion coefficient, $R_s^{\pm}$ denotes the particle radius, while $r$ and $t$ represent radial position and time respectively. The system is subject to initial conditions of uniform concentration:
$c_s^{\pm}(r,0) = c^{\pm}(0)$.
The boundary conditions for Eq.\eqref{eq:1} are:
\begin{equation}
\left. \frac{\partial c_s^{\pm}}{\partial r}(r,t) \right\vert_{r=0} = 0,\label{eq:2}
\end{equation}
at the particle center, and
\begin{equation}
\left. \frac{\partial c_s^{\pm}}{\partial r}(r,t) \right\vert_{r=R_s^{\pm}} = \frac{-j^{\pm} (t)}{D_s^{\pm}},  \label{eq:3}
\end{equation}
at the particle surface, where $j^{\pm}$ represents the lithium-ion flux.

The terminal cell voltage $V_{{\rm SPM}}$ is determined by:
\begin{equation}
\begin{aligned}
V_{{\rm SPM}} (t) = \rm OCP^+(\tilde{c}_{ss}^+(t)) - OCP^-(\tilde{c}_{ss}^-(t)) + \\
\eta^+(\tilde{c}_{ss}^+(t), I(t)) - \eta^-(\tilde{c}_{ss}^-(t), I(t)) - R_0 I(t), 
\end{aligned} 
\label{eq:4}
\end{equation}
where $OCP^{\pm}$ is the open-circuit potential, $\eta^{\pm}$ represents surface overpotential, $I$ denotes applied current, and $R_0$ is the cell's internal resistance. The normalized surface concentration is defined as:
\begin{equation}
\tilde{c}_{ss}^{\pm}(t) = \frac{c_{ss}^{\pm}(t)}{c_{{\rm max}}^{\pm}}, \label{eq:5}
\end{equation}
with $c_{ss}^{\pm}(t) = c_s^{\pm}(R_s^{\pm},t)$ representing the surface concentration.

The surface overpotential is expressed as:
\begin{equation}
\eta^{\pm}(\tilde{c}_{ss}^{\pm}(t), I(t)) = \frac{2RT}{F} \sinh^{-1} \left( \frac{\pm I(t)}{2a^{\pm} L^{\pm} j_{0}^{\pm}(\tilde{c}_{ss}^{\pm}(t))} \right), \label{eq:6}
\end{equation}
where $R$, $T$, and $F$ represent the universal gas constant, temperature, and Faraday's constant respectively. Parameters $a^{\pm}$ and $L^{\pm}$ denote the specific surface area and electrode thickness.

The exchange current density is written as follows:
\begin{equation}
j_{0}^{\pm}(\tilde{c}_{ss}^{\pm}(t)) = r_{{\rm eef}}^{\pm} c_{{\rm max}}^{\pm} \sqrt{c_e \tilde{c}_{ss}^{\pm}(t) (1 - \tilde{c}_{ss}^{\pm}(t))}, \label{eq:7}
\end{equation}
with $r_{{\rm eef}}^{\pm}$ as the electrode reaction rate constant and $c_e$ as the electrolyte lithium-ion concentration, which is assumed as constant in the SPM framework.

The uniform intercalation current density is given by:
\begin{equation}
j^{\pm}(t) = \pm \frac{I(t)}{Fa^{\pm}A^{\pm}L^{\pm}}, \label{eq:8}
\end{equation}
where $A^{\pm}$ represents the electrode surface area.

Eq.~\eqref{eq:1}--\eqref{eq:8} constitute the complete set of governing equations for the SPM model. Upon examination, we observe that the term $\frac{D_s^{\pm}}{(R_s^{\pm})^2}$ appears consistently throughout these equations. This observation leads to an important insight regarding parameter estimation: rather than separately estimating $D_s^{\pm}$ and $R_s^{\pm}$, we can estimate their combined form $\frac{D_s^{\pm}}{(R_s^{\pm})^2}$ as a single parameter. This approach, known as parameter grouping, effectively reduces the number of parameters requiring estimation while maintaining model accuracy. Moreover, each parameter cannot be uniquely identified independently.

Following the parameter grouping methodology presented in \cite{Bizeray2018}, we introduce key parameter groups:
\begin{equation}
\alpha^{\pm} = \frac{(R_s^{\pm})^2}{D_s^{\pm}},\label{eq:9}
\end{equation}
and
\begin{equation}
Q^{\pm} = F A^{\pm} L^{\pm} \varepsilon^{\pm} c_{\text{max}}^{\pm}.\label{eq:10}
\end{equation}

Additionally, we define a grouped kinetic parameter:
\begin{equation}
d^{\pm} = \frac{r_{\text{eef}}^{\pm} \sqrt{c_e}}{F R_s^{\pm}}.\label{eq:11}
\end{equation}

Consequently, the reduced parameter set requiring estimation becomes:
\begin{equation}
\theta = \big[ \alpha^{-}, \alpha^{+}, Q^{-}, Q^{+}, d^{-}, d^{+}, \text{SOC}^{-}_0, \text{SOC}^{+}_0, R_0 \big]^{\top} \in \mathbb{R}^9,
\label{eq:12}
\end{equation}
where \(\text{SOC}^{-}_0 = \text{SOC}^{-}(0)\) and \(\text{SOC}^{+}_0 = \text{SOC}^{+}(0)\) are the initial concentrations of lithium in the solid phase for the positive and negative electrodes, respectively. Through the implementation of parameter grouping methodology, the parameter estimation process is significantly streamlined, requiring only the nine parameters defined in Eq.~\eqref{eq:12}. This reduced parameter set maintains the model's fidelity while substantially decreasing the computational complexity of the estimation procedure.

\subsubsection{Basic Profiles}

In this study, a commercial 18650 lithium-ion battery was employed as the experimental subject. The battery features a nickel manganese cobalt oxide (NMC) cathode and a graphite anode, with a nominal capacity of 2.9 Ah. Parameter identification has been previously conducted on this battery, yielding a set of battery model parameters as presented in Table~\ref{Table1}.The data in Table~\ref{Table1} are derived from our previous study, where parameter estimation was performed using PSO on real battery testing data. For further details, please refer to Reference \cite{guo2025control}. These identified parameters are utilized as the true parameters of the battery model. Based on these parameters and the established battery model, we generated data for five typical operating conditions, obtaining current and voltage values to facilitate subsequent battery parameter identification procedures.

\begin{table}[H]
\centering
\footnotesize
\caption{The model parameters.}
\label{Table1}
\makebox[\textwidth][c]{%
  \begin{tabular}{@{}lll@{}}
    \toprule
    \textbf{Parameter} & \textbf{Value} & \textbf{Unit} \\
    \midrule
    $\alpha^{-}$ & 3105.3457 & 
    $\text{s}$ \\
    $\alpha^{+}$ & 1865.8674 & 
    $\text{s}$ \\
    $Q^{-}$ & 10765.6853 & 
    $\text{C}$ \\
    $Q^{+}$ & 11117.7742 &
    $\text{C}$ \\
    $d^{-}$ & $3.3407 \times 10^{-5}$ & 
    $\text{s} \cdot \text{mol}^{1/2} \cdot \text{m}^{-5/2}$ \\
    $d^{+}$ & $7.3545 \times 10^{-4}$ & 
    $\text{s} \cdot \text{mol}^{1/2} \cdot \text{m}^{-5/2}$ \\
    $\text{SOC}^{-}_0$ & 0.9472 & - \\
    $\text{SOC}^{+}_0$ & 0.0188 & - \\
    $R_0$ & 0.0218 & $\Omega$ \\
    \bottomrule
  \end{tabular}
}
\end{table}

\begin{figure}[H]
\centering
\includegraphics[width=\textwidth]{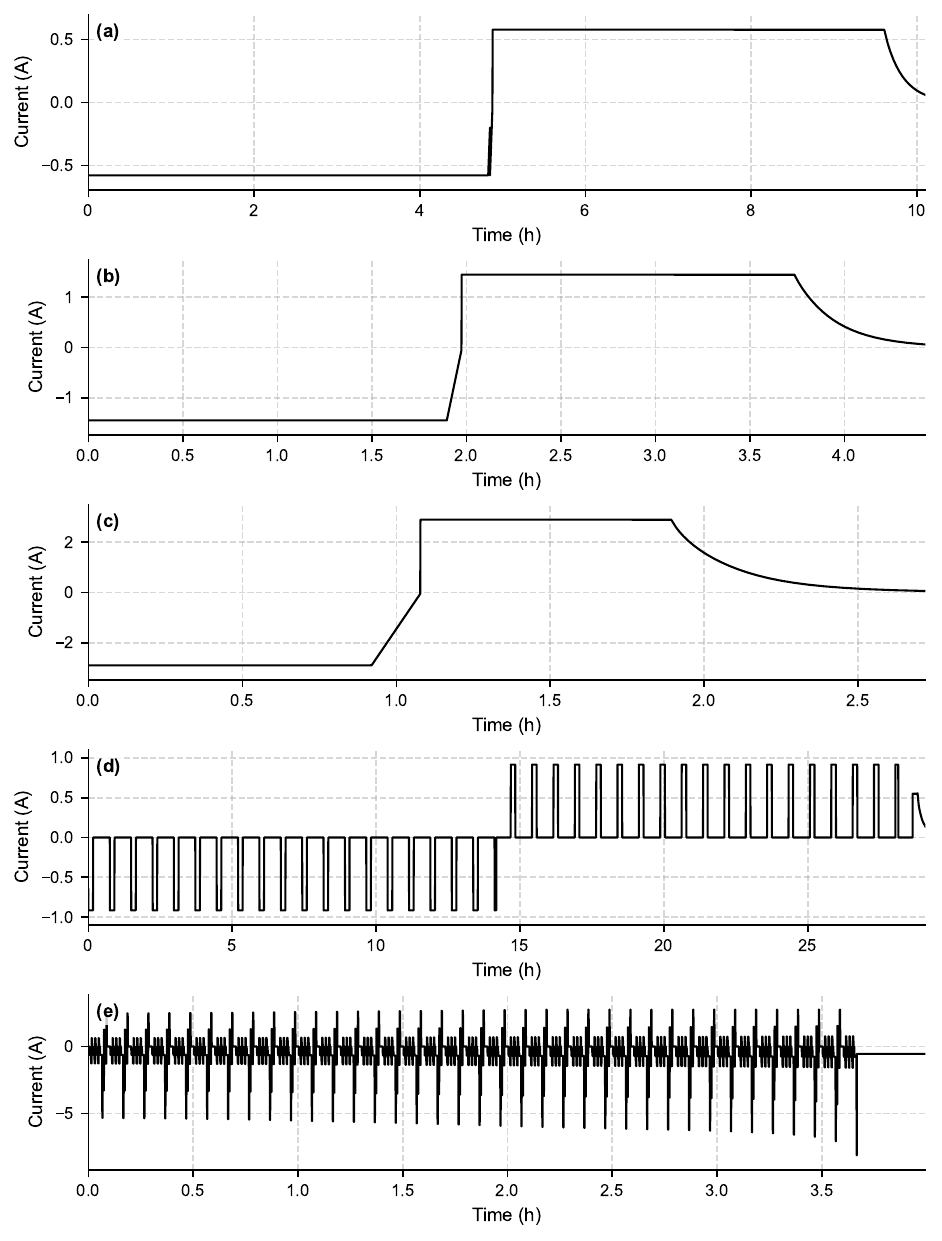}
\caption{Current profiles of the five battery testing protocols considered in this study: (a) constant current discharge at C/5 rate, (b) constant current discharge at C/2 rate, (c) constant current discharge at 1C rate, (d) pulse charging and discharging profile, and (e) DST profile. All current values are shown as functions of time.\label{figure3}}
\end{figure}

Based on the SPM model described in the previous section, five basic operating conditions were generated for battery parameter estimation: CCCV discharge-charge at C/5, C/2, and 1C rates, pulse discharge and charge profiles, and DST. Figure~\ref{figure3} illustrates the current inputs for these five basic operating conditions. The experimental protocols detailed in Table~\ref{Table2} present a comprehensive matrix of battery testing conditions. Since the SPM neglects concentration variations in the electrolyte, its prediction errors typically increase under high current conditions, generally above 1C \cite{moura2016battery}. As shown in the second column of Table 2, the maximum equivalent discharge rate across all five operating profiles is C/1.4, which is below 1C. Although the DST profile includes brief 2C discharge pulses, it does not involve long-duration 2C discharges. Therefore, the operating conditions considered in this study remain consistent with the applicability of the SPM.

\begin{table}[H]
\caption{Characterization of battery testing protocols with operating parameters\label{Table2}}
\label{results1}
\centering
\resizebox{\textwidth}{!}{ 
\begin{tabular}{l c c c c}
\toprule
Protocol Description & CC equivalent C-rate[C] & Duration [h] & Maximum Current [C]* & Designation \\
\midrule
CCCV (C/5) & C/5.2 & 10.1 & C/5 & C/5 \\
CCCV (C/2) & C/2.3 & 4.4 & C/2 & C/2 \\
CCCV (1C) & C/1.4 & 2.7 & 1C & 1C \\
Pulse (9.5-min C/3; 35-min rest) & C/14.6 & 29.1 & C/3 & P \\
DST (2C, 1C, C/2, C/5) & C/3 & 4 & [1C, 2C] & DST \\
\bottomrule
\multicolumn{5}{l}{\footnotesize{* For protocols with differing maximum charge ($\bar{I}_{\rm ch}$) and discharge ($\bar{I}_{\rm dch}$) currents, values are presented as an interval [$\bar{I}_{\rm ch}$, $\bar{I}_{\rm dch}$].}}
\end{tabular}
}
\end{table}

The experimental investigation begins with three fundamental CCCV discharge-charge protocols, each implementing a complete cycle of CCCV discharge followed by CCCV charge at different C-rates. In each protocol, the battery first undergoes a CCCV discharge process where a constant current is applied until reaching the cut-off voltage, followed by a constant voltage phase until the current drops below a predetermined threshold. Subsequently, a CCCV charging process is conducted, applying a constant current until reaching the upper voltage limit, followed by a constant voltage charging phase until the current decreases to the specified cut-off value. For the constant voltage (CV) phase of CCCV operations, a proportional controller was implemented. The CV phase was terminated when the current magnitude decreased below 50mA, which can be expressed as:
\begin{equation}
|I_{CV}| < 50 \text{ mA} \rightarrow \text{End of CV phase}.\label{eq:13}
\end{equation}

The first protocol operates at C/5 rate, where both discharge and charge processes maintain a maximum current of C/5. This cycling approach requires approximately 10.1 hours for completion, allowing thorough investigation of battery behavior under low-stress conditions. The second protocol elevates the rate to C/2, completing the full discharge-charge cycle within 4.4 hours. The third protocol further intensifies the process to 1C, where both discharge and charge operate at 1C rate, completing the cycle within 2.7 hours. This accelerated cycling enables investigation of battery performance under more demanding conditions. The CC equivalent C-rate refers to the current intensity under equivalent constant-current conditions. We calculate this using the mean absolute value of current intensity. Due to the presence of the CV phase, the CC equivalent C-rate is lower than the current in their respective constant current conditions. For nominal rates of C/5, C/2, and 1C, the corresponding CC equivalent C-rates are C/5.2, C/2.3, and C/1.4, respectively.These systematically designed CCCV protocols, incorporating both constant current and constant voltage phases in both discharge and charge processes, provide comprehensive insights into battery behavior across different operating rates, from cycling at C/5 to more aggressive conditions at 1C.

The fourth protocol introduces a more sophisticated pulse discharge-charge methodology, denoted as P in Table~\ref{Table2}. This regime implements alternating phases of C/3 current pulses, each lasting one minute, interspersed with five-minute rest periods. The intermittent nature of this profile results in an effective C-rate equivalent to C/14.6, extending over a substantial duration of 29.1 hours while maintaining a maximum current of C/3 during active phases. This pulsed approach enables detailed examination of battery recovery characteristics during rest periods and the impact of interrupted charging/discharging cycles on overall battery performance.

The fifth protocol employs a DST profile, incorporating a complex sequence of varying current rates (2C, 1C, C/2, and C/5) in a cyclic pattern. This dynamic profile achieves an equivalent C-rate of C/3 and completes within 4 hours. The maximum current values for discharging ang charging are 2C and 1C.

\subsection{Test Matrix Development}

In this section, we generate multiple composite operating conditions by combining the five basic conditions obtained in the previous section. The CCCV and pulse profiles inherently contain both discharge and charge processes, with each cycle beginning and ending at a fully charged state (SOC = 1). This characteristic allows these profiles to be directly connected in sequence without introducing inconsistencies in the SOC. A notable consideration pertains to the DST profile, which exclusively comprises discharge operations starting from SOC = 1 and terminating at SOC = 0. Furthermore, the DST protocol is specifically designed to initiate from a fully charged battery state. For this reason, the DST must be positioned as the terminal segment in any composite sequence to ensure that the required initial conditions are satisfied and that no subsequent profiles are compromised by its terminating at a depleted state.

It is also important to clarify that the sequence of the operating conditions does not affect the parameter estimation results in our study. While the DST must remain last, all other profiles begin and end at SOC = 1, which guarantees that the starting point of the following profile remains unaffected regardless of the order in which they are arranged. Consequently, these profiles can be reordered freely without influencing the subsequent states of charge. Moreover, in the PSO-based optimization framework adopted in this work, the loss function is defined as the root-mean-square error (RMSE) over the entire voltage sequence. Because all non-DST profiles consistently return to SOC = 1, reordering them does not alter the overall RMSE calculation and therefore does not influence the optimization process.

The test matrix presented in Table~\ref{Table3} establishes a comprehensive battery evaluation protocol comprising 31 distinct test scenarios. This systematic testing framework integrates CCCV cycling at multiple rates with dynamic stress testing to thoroughly assess battery performance across diverse operational conditions. The testing matrix employs a hierarchical structure that progresses from complex multi-test configurations to fundamental single tests, enabling detailed analysis of both individual and combined effects of various testing. For the combination and ordering of the five operating conditions, we prioritize CCCV conditions first, sorting them in ascending order of current. Then, we include dynamic conditions, with the DST condition placed at the end if applicable. We disregard the impact of different condition sequences on the results, as actual battery testing typically follows this logic, aligning more closely with real laboratory testing procedures.

The test scenarios are organized in a hierarchical structure based on the number of tests evaluated simultaneously. Scenario 1 represents the most comprehensive configuration, incorporating all five conditions: three CCCV rates (C/5, C/2, 1C) combined with both dynamic test types (Pulse Testing and DST). This complete configuration enables the most thorough evaluation of test interactions.

Four-condition configurations are examined in Scenarios 2-6. These Scenarios systematically remove one test from the complete set, creating various combinations such as triple-rate CCCV with single dynamic testing (Scenarios 2-3) and dual-rate CCCV with both dynamic tests (Scenarios 4-6). This approach allows for detailed analysis of how the absence of specific tests affects overall battery performance.

Scenarios 7-16 investigate three-condition combinations, incorporating various arrangements of CCCV rates and dynamic tests. These configurations include triple-rate CCCV, dual-rate CCCV with single dynamic testing and single-rate CCCV with both dynamic tests, providing insights into the interactions between smaller tests.

Two-condition combinations are explored in Scenarios 17-26, examining paired CCCV rates, CCCV rate with dynamic test combinations, and the interaction between both types of dynamic testing. These simplified configurations enable clear identification of test pair effects on battery behavior.

The final segment, Scenarios 27-31, evaluates single-condition configurations, establishing baseline performance data for each individual CCCV rate (C/5, C/2, 1C) and dynamic test type (Pulse Testing, DST). These fundamental Scenarios provide essential reference points for quantifying the effects of test combinations in more complex configurations.

This methodically structured testing matrix offers several significant advantages. The systematic test grouping facilitates comprehensive analysis of both individual and combined test effects, while the hierarchical organization enables clear identification of interaction effects by comparing results across different test combinations.

\begin{table}[H]
\centering
\footnotesize
\caption{Summary of test Scenarios with different protocol combinations and experiment duration\label{Table3}}
\makebox[\textwidth][c]{%
  \begin{tabularx}{\textwidth}{@{} >{\centering\arraybackslash}p{4cm} >{\raggedright\arraybackslash}X >{\centering\arraybackslash}p{5cm} @{}}
    \toprule
    \textbf{Scenario No.} & \textbf{Test Configuration}\textsuperscript{1} & \textbf{Experiment Duration [h]} \\
    \midrule
    1  & C/5, C/2, 1C, P, DST  & 50.3 \\
    \midrule
    2  & C/5, C/2, 1C, P       & 46.3 \\
    3  & C/5, C/2, 1C, DST     & 21.2 \\
    4  & C/5, C/2, P, DST      & 47.6 \\
    5  & C/5, 1C, P, DST       & 45.9 \\
    6  & C/2, 1C, P, DST       & 40.2 \\
    \midrule
    7  & C/5, C/2, 1C          & 17.2 \\
    8  & C/5, C/2, P           & 43.6 \\
    9  & C/5, C/2, DST         & 18.5 \\
    10 & C/5, 1C, P            & 41.9 \\
    11 & C/5, 1C, DST          & 16.8 \\
    12 & C/5, P, DST           & 43.2 \\
    13 & C/2, 1C, P            & 36.2 \\
    14 & C/2, 1C, DST          & 11.1 \\
    15 & C/2, P, DST           & 37.5 \\
    16 & 1C, P, DST            & 35.8 \\
    \midrule
    17 & C/5, C/2              & 14.5 \\
    18 & C/5, 1C               & 12.8 \\
    19 & C/5, P                & 39.2 \\
    20 & C/5, DST              & 14.1 \\
    21 & C/2, 1C               & 7.1 \\
    22 & C/2, P                & 33.5 \\
    23 & C/2, DST              & 8.4 \\
    24 & 1C, P                 & 31.8 \\
    25 & 1C, DST               & 6.7 \\
    26 & P, DST                & 33.1 \\
    \midrule
    27 & C/5                   & 10.1 \\
    28 & C/2                   & 4.4 \\
    29 & 1C                    & 2.7 \\
    30 & P                     & 29.1 \\
    31 & DST                   & 4.0 \\
    \bottomrule
  \end{tabularx}%
}
\noindent{\footnotesize{\textsuperscript{1} P - Pulse test (29.1h); DST - Dynamic Stress Test (4.0h); C/5 (10.1h), C/2 (4.4h), 1C (2.7h) - CCCV discharge-charge rates}}
\end{table}

\subsection{Parameter Estimation and Evaluation}
The 31 operating conditions derived in the previous section serve as input profiles for parameter estimation. 
In this study, the PSO algorithm is employed for parameter identification. 
PSO has been widely applied to battery parameter estimation and has demonstrated strong global search capability, accuracy, and stability in our previous work \cite{Guo2024,Guo2024b}. 
Therefore, it is adopted here as the optimization method for parameter estimation of the battery model. All computational implementations in this study were developed in Python programming language. The PSO algorithm was implemented using the 'pyswarms' library (version 1.3.0). Parameter estimation was performed on a computing system equipped with a 16-core CPU processor, 32GB of RAM, running a Linux operating system. The optimization process can be expressed mathematically as:
\begin{equation}
\theta^* = \arg\min_{\theta} \sum_{k=1}^{N} (V_{\text{measured}, k} - V_{\text{model}, k}(\theta))^2,\label{eq:14}
\end{equation}
where $\theta^*$ represents the optimal parameter set, $V_{\text{measured},k}$ denotes the measured voltage at time step $k$, and $V_{\text{model}, k}(\theta)$ represents the model-predicted voltage with parameter set $\theta$.

To evaluate the accuracy of each parameter set, we employ a comprehensive validation approach using the RMSE metric. Our validation approach systematically assessed each of the 31 parameter sets by simulating all 31 operating conditions and calculating their respective RMSE values. This generated a total of 961 validation test (31 parameter sets × 31 operating conditions), providing a thorough assessment of how each parameter set performed across the complete range of operational conditions. This extensive cross-validation methodology enabled us to rigorously evaluate the robustness and generalization capability of each parameter set across diverse operational Scenarios. The RMSE for each parameter set is calculated as:
\begin{equation}
\mathrm{RMSE}_i = \sqrt{\frac{1}{N}\sum_{k=1}^{N} (V_{\text{measured},k} - V_{\text{model},k}(\theta _i^*))^2},\label{eq:15}
\end{equation}
where ${\rm RMSE}_i$ represents the error metric for the parameter set obtained from the $i$-th operating condition, and $N$ is the total number of data points in the complete combination profile.

Since 31 operating conditions were generated using the SPM, we have access to the true parameter values that were used to create these synthetic profiles. This provides a unique advantage for validation, as we can directly compare the estimated parameters against these known true values. The true value can be found in Table~\ref{Table1}. Consequently, we calculate the percentage error between the estimated parameters and the true parameters used to generate the synthetic data:
\begin{equation}
\delta_{\theta,i} = \frac{|\theta_i^* - \theta_{\rm true}|}{\theta_{\rm true}} \times 100\% ,\label{eq:16}
\end{equation}
where $\delta_{\theta,i}$ represents the parameter percentage error for the $i$-th operating condition. This $\delta_{\theta,i}$ is a vector containing the percentage errors for each parameter within this parameter set.

Besides this component-to-component error, we are also interested in the distance between the nominal and estimated parameter vectors, for which we use the Euclidean distance $\delta_{\rm dist}$ between the two vectors $\theta^*$ and $\theta_{\rm true}$ for the $i$-th operating conditions defined as
\begin{equation}
\delta_{{\rm dist},i} =\lVert \theta_i^* - \theta_{\rm true} \lVert_2 = \sqrt{ (\theta^*_i - \theta_{\rm true})^\top (\theta^*_i - \theta_{\rm true}) }.
\label{delta}
\end{equation}

A primary objective of this investigation is to analyze the impact of different input conditions on parameter identification computational efficiency. The maximum iteration count in PSO serves as a critical determinant of algorithmic runtime, which can be represented as: 
\begin{equation}
T_{\rm opt} = N_{\rm iterations} \times T_{\rm per \; iteration},\label{eq:17}
\end{equation}
where$T_{\rm opt}$ is the total computational time of the optimization problem and $T_{\rm per \; iteration}$ represents the time required per iteration.
While increasing the maximum iteration limit theoretically enhances estimation accuracy, it comes with a proportional increase in computational overhead. To establish an optimal balance between estimation accuracy and computational efficiency, this study implements a uniform maximum iteration limit of $N_{\rm iterations} = 300$ iterations as the termination criterion, consistent with previous studies on electrochemical model parameter identification in the literature \cite{Fan2020}. Another relevant temporal metric to be considered as a burden is the time required to perform the experiments themselves, which is here denoted as $T_{\rm exp}$, and it depends on the tested profile(s). That is the total duration in Table\ref{Table3}. By accounting for these two time requirements, the total time $T_{\rm total}$ is computed as:
\begin{equation}
T_{\rm total} = T_{\rm opt} + T_{\rm exp}.\label{eq:17b}
\end{equation}

The research methodology involves executing the PSO algorithm independently for each of the thirty-one input conditions, with comprehensive documentation of computational runtime, parameter percentage error results, and validation RMSE for each Case. The performance metrics for each run can be represented as:
\begin{equation}
T_{{\rm total},i}, \; \delta_{\theta,i}, \; {\rm RMSE}_i \; \text{ for } \; i = 1,2,\ldots,31,\label{eq:18}
\end{equation}
where $T_{{\rm total},i}$ represents the computational time, $\delta_{\theta,i}$ denotes parameter percentage error, and ${\rm RMSE}_i$ indicates the validation error for the $i$-th input condition.By examining the trade-offs between these metrics across all 31 operating conditions, we can determine which condition or set of conditions yields the most favorable combination of rapid convergence and accuracy parameter estimates.

\section{Results \& Discussion}

\begin{figure}[H]
\centering

\makebox[\textwidth][c]{%
    \includegraphics[width=1.4\textwidth,trim={2cm 0.5cm 2cm 1cm},clip]{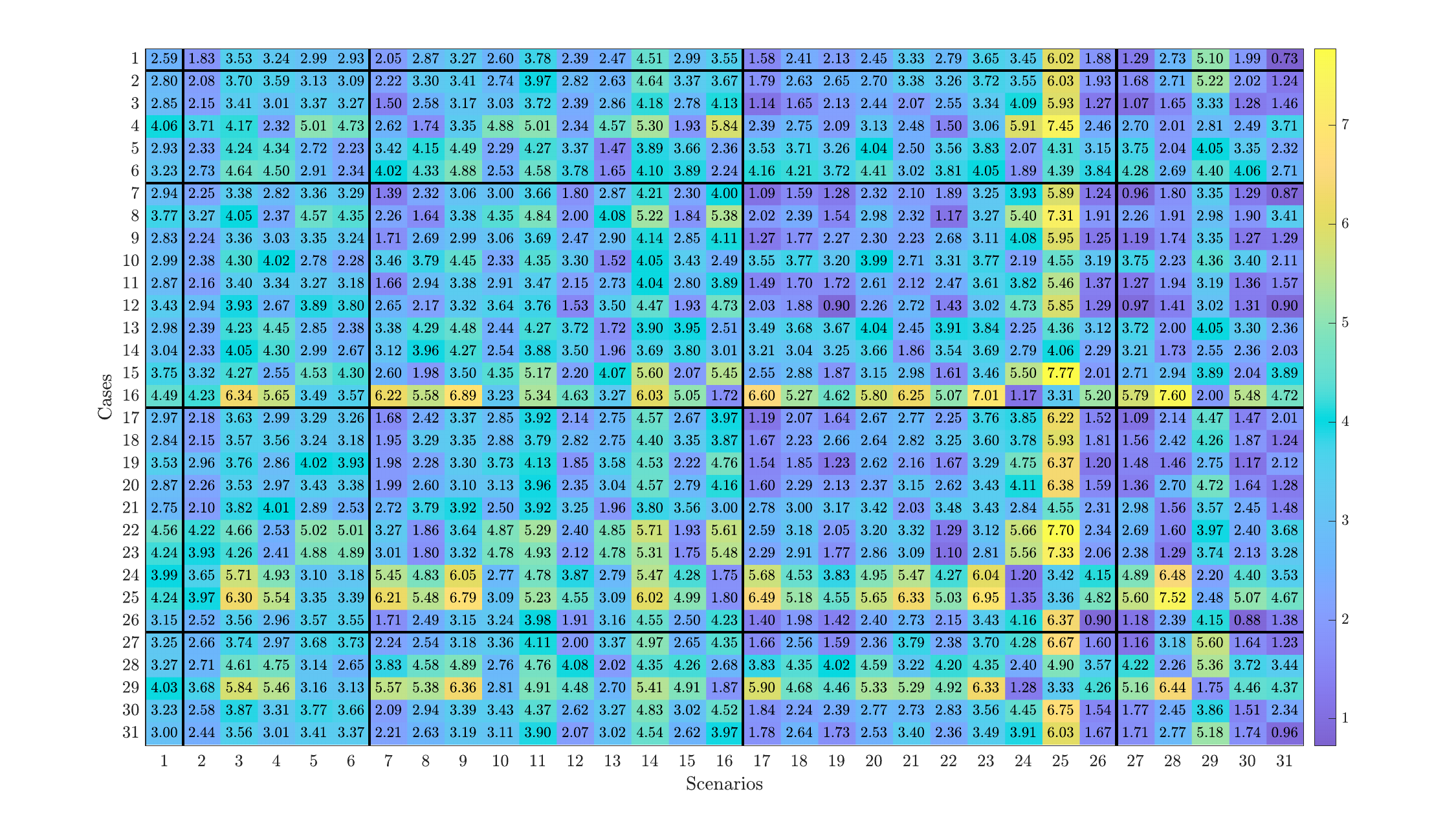}
}
\caption{RMSE matrix obtained for each considered Case and Scenario.
The reported values are $10^2\times$RMSE.}
\label{figure4}
\end{figure}

The obtained results are now presented in terms of different simulations and metrics. First, we show the case where simulations were performed for each of the five considered charge/discharge profiles (i.e. C/5, C/2, 1C, P and DST) as well as all possible combinations of these profiles, giving rise to the 31 profiles tested here. 
First, each of these 31 profiles are used in an optimization problem to find the optimal parameters for that particular profile by using a global minimizer as explained above. 
Thus, 31 optimal parameter sets $\theta^*$ in Eq.\eqref{eq:14} are obtained, one per used profile, which is denoted as Cases. 
Each of these parameter vectors is then evaluated in terms of output voltage error, the RMSE in Eq.\eqref{eq:15}, for all the 31 considered profiles, which is denoted as Scenarios. A total of 961 validation test (31 parameter sets × 31 operating conditions) can be obtained.
This yields the 31$\times$31 matrix shown in Figure \ref{figure4} whose x-axis corresponds to the Scenarios (31 operating conditions) and the y-axis corresponds to the Cases (31 parameter sets). Each row of this matrix reflects a different parameter estimate for each considered profile, and each column is the model evaluation of that particular parameter vector considering the specific profile. 

The obtained results are analyzed in the following levels (from L0 to L7):
\begin{enumerate}
    \item[L0] validation by Scenario 1;
    \item[L1] overall minimum RMSE value;
    \item[L2] minimum RMSE value in the diagonal elements;
    \item[L3] minimum value obtained for each column;
    \item[L4] second smallest value obtained for each column;
    \item[L5] maximum value obtained for each column;
    \item[L6] parameter errors and time requirements;
    \item[L7] overall best datasets for given performance metrics.
\end{enumerate}
By looking at these levels we will be able to understand the implications of the considered situations.

\subsection{L0: Validation by Scenario 1}

\begin{figure}[H]
\centering

\makebox[\textwidth][c]{%
    \includegraphics[width=1.4\textwidth]{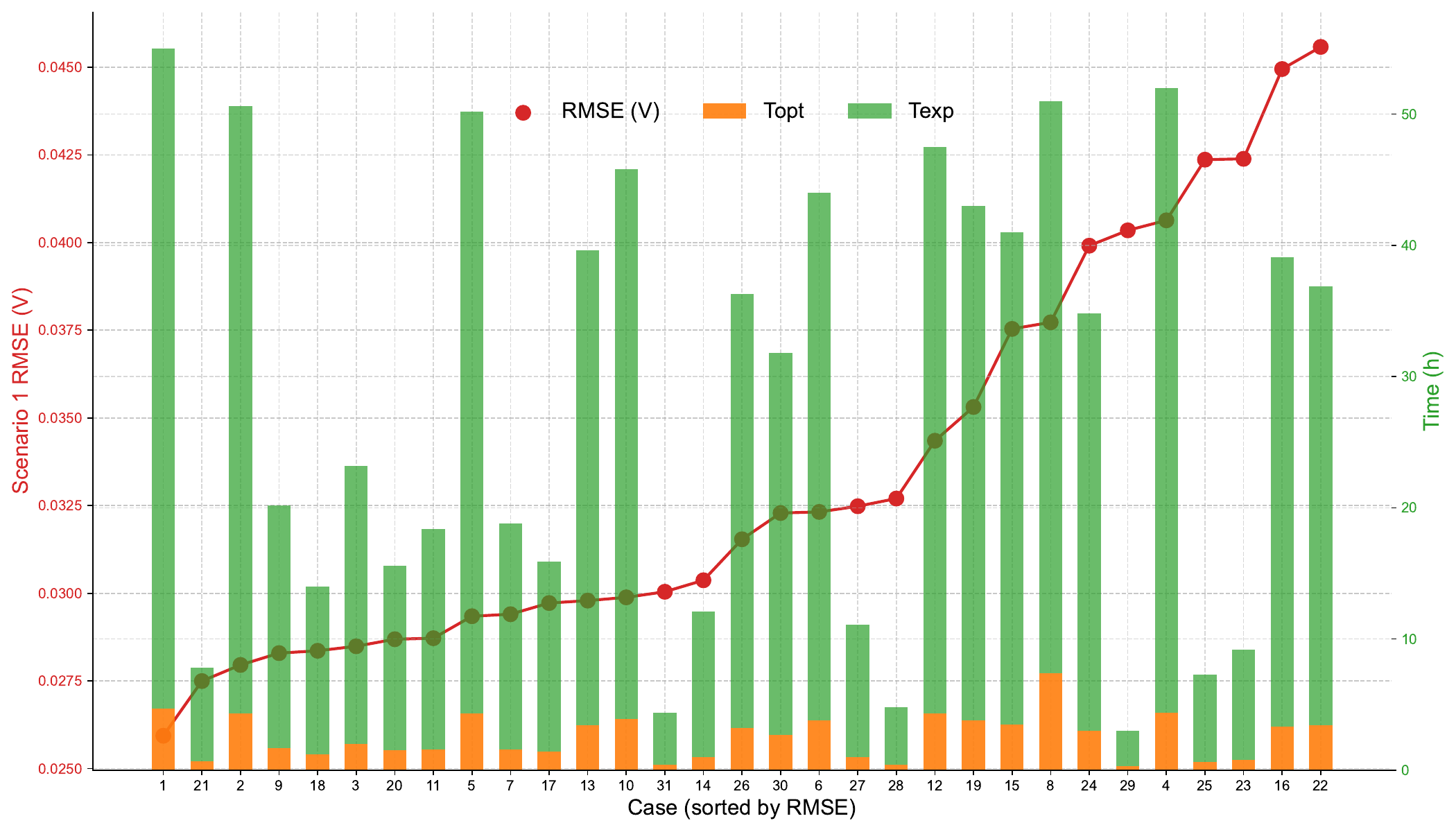}
}
\caption{Scenario 1 RMSE vs. Total Time by Cases(parameter sets).}
\label{figure5}
\end{figure}

Due to the comprehensive nature of Scenario 1, which encompasses all five fundamental operating conditions, we initially employed this Scenario to evaluate the complete set of 31 battery parameter combinations (Cases). The results illustrated in Figure~\ref{figure5} demonstrate that Case 1\{C/5,C/2,1C,P,DST\} exhibited the minimal RMSE. This superior performance can be attributed to the fact that Case 1 was specifically derived through estimation using Scenario 1's conditions, thus naturally achieving optimal results within this Scenario.

Analysis of both Figure~\ref{figure5} and Table~\ref{table4} reveals that Case 1 required the highest total time among all Cases. The total time is calculated by Eq.~\eqref{eq:17b}. This extended duration is a direct consequence of its comprehensive inclusion of all five operating conditions, resulting in prolonged experimental procedures and parameter estimation processes. A particularly noteworthy observation is that Case 21\{C/2,1C\}'s parameter estimation error was second only to Case 1, while requiring merely 14\% of Case 1's total time. In other words, Case 21 achieved parameter estimation approximately seven times faster than Case 1.
\begin{table}[H]
\centering
\footnotesize
\caption{Performance metrics for each Case.\label{table4}}
\makebox[\textwidth][c]{%
  \begin{tabularx}{\textwidth}{@{}Xcccc@{}}
    \toprule
    \textbf{Case} & \textbf{T\textsubscript{opt}[h]} & \textbf{T\textsubscript{exp}[h]} & \textbf{Total [h]} & \textbf{Scenario 1 RMSE (V)} \\
    \midrule
    1  & 4.7 & 50.3 & 55.0 & 0.0259 \\
    2  & 4.3 & 46.3 & 50.6 & 0.0280 \\
    3  & 2.0 & 21.2 & 23.2 & 0.0285 \\
    4  & 4.4 & 47.6 & 52.0 & 0.0406 \\
    5  & 4.3 & 45.9 & 50.2 & 0.0293 \\
    6  & 3.8 & 40.2 & 44.0 & 0.0323 \\
    7  & 1.6 & 17.2 & 18.8 & 0.0294 \\
    8  & 7.4 & 43.6 & 51.0 & 0.0377 \\
    9  & 1.7 & 18.5 & 20.2 & 0.0283 \\
    10 & 3.9 & 41.9 & 45.8 & 0.0299 \\
    11 & 1.6 & 16.8 & 18.4 & 0.0287 \\
    12 & 4.3 & 43.2 & 47.5 & 0.0343 \\
    13 & 3.4 & 36.2 & 39.6 & 0.0298 \\
    14 & 1.0 & 11.1 & 12.1 & 0.0304 \\
    15 & 3.5 & 37.5 & 41.0 & 0.0375 \\
    16 & 3.3 & 35.8 & 39.1 & 0.0449 \\
    17 & 1.4 & 14.5 & 15.9 & 0.0297 \\
    18 & 1.2 & 12.8 & 14.0 & 0.0284 \\
    19 & 3.8 & 39.2 & 43.0 & 0.0353 \\
    20 & 1.5 & 14.1 & 15.6 & 0.0287 \\
    21 & 0.7 & 7.1  & 7.8  & 0.0275 \\
    22 & 3.4 & 33.5 & 36.9 & 0.0456 \\
    23 & 0.8 & 8.4  & 9.2  & 0.0424 \\
    24 & 3.0 & 31.8 & 34.8 & 0.0399 \\
    25 & 0.6 & 6.7  & 7.3  & 0.0424 \\
    26 & 3.2 & 33.1 & 36.3 & 0.0315 \\
    27 & 1.0 & 10.1 & 11.1 & 0.0325 \\
    28 & 0.4 & 4.4  & 4.8  & 0.0327 \\
    29 & 0.3 & 2.7  & 3.0  & 0.0403 \\
    30 & 2.7 & 29.1 & 31.8 & 0.0323 \\
    31 & 0.4 & 4.0  & 4.4  & 0.0300 \\
    \bottomrule
  \end{tabularx}%
}
\end{table}

These findings suggest that Case 21 represents a compromise between accuracy and efficiency in parameter estimation. It maintains a high level of estimation accuracy while significantly reducing the computational burden. This makes Case 21 an extremely attractive option for practical applications where both precision and time efficiency are crucial considerations. The substantial reduction in computational time, coupled with only a marginal decrease in accuracy compared to Case 1, positions Case 21 as a highly efficient alternative for battery parameter estimation in real-world applications.

\subsection{L1: Overall Minimum RMSE Value}

The L1 level is found by inspecting the minimum value(0.0073V) of the RMSE corresponding to the position (1,31) in the matrix using the (Case,Scenario) notation. In other words, use the first parameter set obtained from\{C/5, C/2, 1C, P, DST\} and verify it with the 31st operating condition \{DST\}.
This tells us that the overall minimum error is obtained when the complete dataset is used to estimate the model parameters and only the DST data is used for model validation. Thus, the error obtained by considering the full dataset is larger than the one resulting from using DST alone (position (1,1) in the matrix), which implies that the error associated to the complementary part of the DST data in the full dataset (where full dataset is \{C/5, C/2, 1C, P, DST\} and the complement of the DST in the full dataset is \{C/5, C/2, 1C, P\}) involves a larger error than the DST itself. This statement is verified through the following derivation
\begin{eqnarray}
{\rm RMSE}_{\rm DST} &\!\!\!\!<&\!\!\!\! {\rm RMSE}_{\rm full}, \label{rmsel1} \\
\sqrt{\frac{{e}^2_{\rm DST}}{N_{\rm DST}}} &\!\!\!\!<&\!\!\!\! \sqrt{\frac{{e}^2_{\rm DST}+{e}^2_{\rm comp}}{N_{\rm DST}+N_{\rm comp}}}, \label{rmsel1b} \\
N_{\rm DST} {e}^2_{\rm DST} + N_{\rm comp} {e}^2_{\rm DST} &\!\!\!\!<&\!\!\!\! N_{\rm DST} {e}^2_{\rm DST} + N_{\rm DST} {e}^2_{\rm comp}, \label{rmsel1c} \\
\frac{{e}^2_{\rm DST}}{N_{\rm DST}} &\!\!\!\!<&\!\!\!\! \frac{{e}^2_{\rm comp}}{N_{\rm comp}}, \label{rmsel1d} \\
{\rm RMSE}_{\rm DST} &\!\!\!\!<&\!\!\!\! {\rm RMSE}_{\rm comp}, \label{rmsel1e}
\end{eqnarray}
where the subscripts ${\rm DST}$, ${\rm full}$ and ${\rm comp}$ denote the datasets used during validation (for the given full dataset used for estimation), namely DST dataset, full dataset and complementary part of the DST data in the full dataset, respectively, 
${e}^2$ denotes the squared error defined as ${e}^2 = e^\top e$ with $e = [e_1, e_2, \ldots, e_N]^\top$ and 
$e_k = V_{\text{measured},k} - V_{\text{model},k}(\theta^*), \; k \in \{1,\ldots,N\}$, as indicated in \eqref{eq:15} and for a data length $N$. 
Indeed, Eq.~\eqref{rmsel1} is verified by inspecting Figure \ref{figure4} where entry (1,31) is ${\rm RMSE}_{\rm DST} = 0.0073$ V and 
entry (1,1) is ${\rm RMSE}_{\rm full} = 0.0259$ V due to the fact that Eq.~\eqref{rmsel1e} also occurs with entry (1,2) as ${\rm RMSE}_{\rm comp} = 0.0183$ V.

Interestingly enough, when the DST is used to estimate parameters, the error is actually larger than this one, see entry (31,31) in Figure~\ref{figure4} where ${\rm RMSE}_{\rm DST}^{\rm DST} = 0.0096 > {\rm RMSE}_{\rm DST}^{\rm full} = 0.0073$. 
Here, superscripts denote the data used for estimation while subscripts denote the validation data to be able to distinguish between both situations. 
This error difference is relatively small, since it is the fourth smallest error of the 31 reported ones in column 31. This discrepancy is likely due to the fact that the parameter estimator based on the PSO algorithm have not fully converged after the specified number of iterations $N_{\rm iter} = 300$, but found a minimum for the DST Case that is slightly larger to the one corresponding to the full dataset. Notice that the selected $N_{\rm iter}$ allowed us to run the optimization problems in a reasonable amount of time and we indeed verified that smaller errors could be obtained if this value was increased (e.g. $N_{\rm iter} = 1000$) at the expense of a sizable amount of computation time required.
On the other hand, we also specified an $N_{\rm iter}$ consistent with the literature and we wanted to show how such a choice could impact the optimization results. This is often overlooked in the literature where only experimental data is directly considered to parameterize the model without any further thoughts or special care.

\subsection{L2: Minimum RMSE Value in The Diagonal Elements}
The L2 condition appears in the position (26,26) of the matrix in Figure~\ref{figure4}, corresponding to the situation where model parameters are estimated and evaluated considering the more dynamic profiles of pulses and DST. Moreover, this error is also minimum for column 26, meaning that the use of this profile for parameter estimation indeed yields the minimum error at the validation stage as well. Even if this yields the minimum error when validating the estimated parameters in the same dataset, this does not correspond to the minimum error that can be obtained throughout validation sets which will be explored in condition L6 below.

\subsection{L3: Minimum Value Obtained for Each Column}
\begin{figure}[H]
\centering

\makebox[\textwidth][c]{%
    \includegraphics[width=1.4\textwidth,trim={2cm 0.5cm 2cm 1cm},clip]{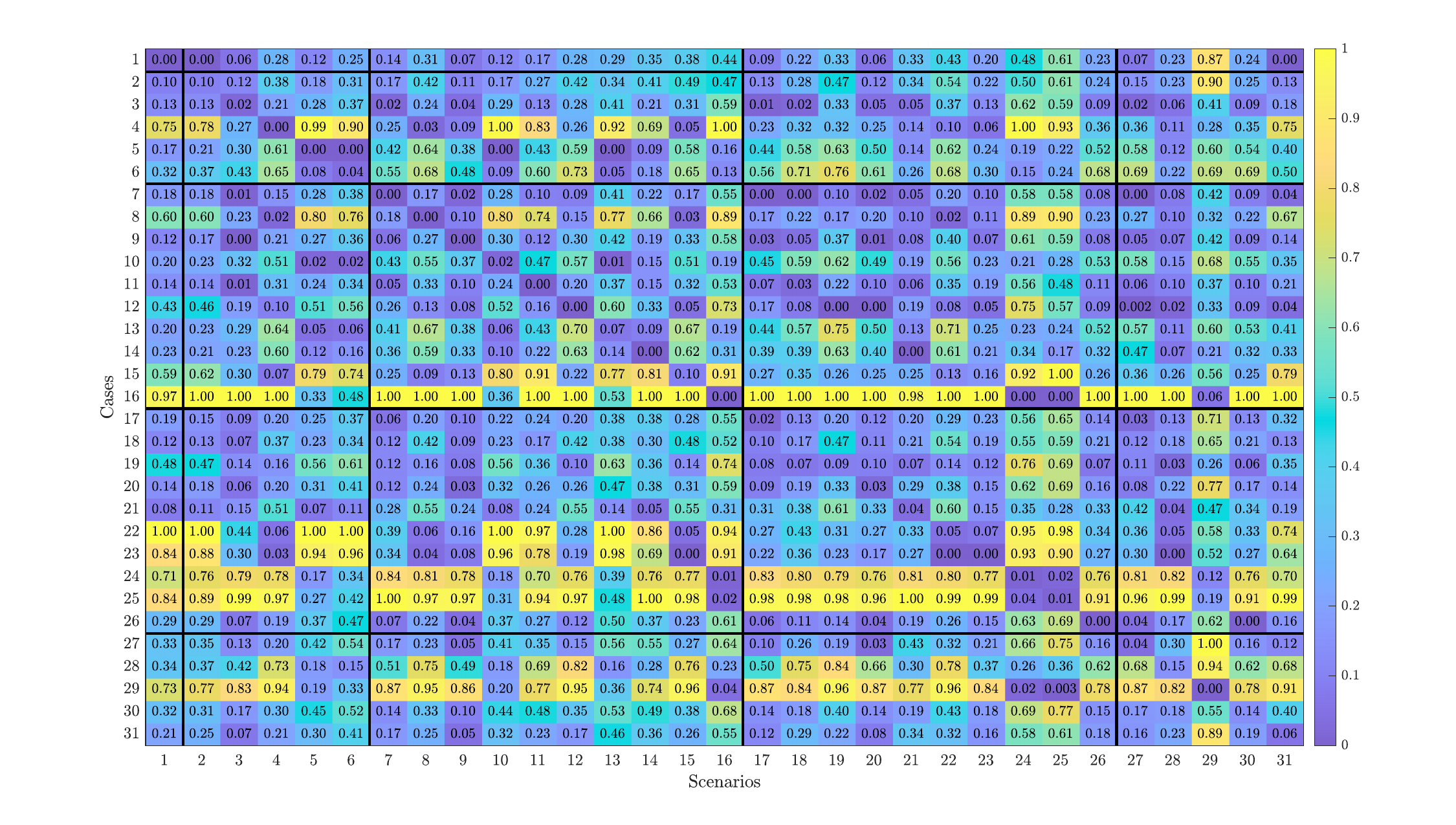}
}
\caption{Column-wise normalized version of RMSE matrix in Figure~\ref{figure4} to stress minimum and maximum values.}
\label{figure6}
\end{figure}

Let us now see the results of the more specific situations associated to L3. 
In order to do so, we transform the RMSE data of Figure~\ref{figure4} into a column-normalized version of this data shown in Figure~\ref{figure6} where the RMSE values have been scaled between minimum and maximum values for a given column, i.e. the minimum and maximum values of a column now correspond to 0 and 1, respectively. 
In the ideal situation, we would expect that the minimum errors appear in the diagonal of the matrix, since this reflects the situations where the parameters are estimated with a given profile and validation is done with that same profile resulting in minimal error. 
From Figure~\ref{figure6}, it can be observed that smaller errors are generally concentrated along the diagonal, which is as expected. However, there are some exceptions where the errors are not the smallest. 
In the following, we explore the particular situations.

First, notice that 13 situations out of the 31 assessed ones actually show the minimum value of the column corresponding to the Case where the model validation error is minimal for the same profile used to estimate the model parameters. 
Since both estimation and validation data are equivalent, datasets do not need to be added nor excluded nor swapped. 
This first group of Scenarios are \{1,4,5,7,8,9,11,12,14,16,23,26,29\} as shown in Table \ref{Table5}. 
In a second group of Scenarios, datasets are expanded in order to include additional profiles to achieve a minimum value during validation, which corresponds to 14 Scenarios (see Table \ref{Table5}). 
From these Scenarios, 8 Scenarios consistently include the DST profile while other 3 Scenarios add the P profile, which highlights the use of more dynamic data to reduce the RMSE during validation. 
The remaining 3 Scenarios include profiles like C/2 and 1C (Scenarios \{17,18,27\} in the table) going from 1 or 2 datasets to 3 datasets. 
The third group of Scenarios involve the exclusion of datasets to reach minimum RMSE for validation, which are more limited than the previous Scenarios with only 2 instances (Scenarios \{3,15\} in the Table~\ref{Table5}). 
In this situations, 1C data and P data were not considered in the validation. 
Finally, the last group of Scenarios corresponds to swaps of datasets, which took place 3 times (Scenarios \{6,13,22\} in the table) where C/2 and P data were replaced by C/5 and DST data, respectively.

There are multiple possible reasons for these changes in the data structure in order to reach minimum output errors during validation. 
The most popular situation where the addition of the DST to the dataset results in the minimum RMSE can be explained by the fact that the inclusion of more dynamically rich data as the DST allows the parameterized model to be more flexible and have better generalization ability to handle other more subtle Cases. In contrast, the estimation without DST in the simpler dataset might yield a model that is too "stiff" or overfits to that specific (more steady-state or constant operating conditions) data and therefore it fails to generalize by not capturing transient and dynamic behaviour accurately. 
This overfitting can lead to better performance only on simpler Cases, not on a broader profiles when evaluated, making the other parameters more robust even when evaluated with another profile. Similar reasoning can be applied when P or other CC profiles are added to minimize validation RMSE since they enrich the dataset and generalize the model. 

The other situations are more limited. 
In the exclusion situation, only 1C and P data are excluded from their respective datasets (Scenarios 3 and 15), which might indicate the introduction of bias toward the behaviour of these profiles, reducing the model ability to handle the dominant dynamic changes of the data effectively. Indeed, 1C is one of the largest currents in the considered datasets sustained for the longest period of time while P is the longest considered profile (see Table \ref{Table5}). 

In the swapping situation, C/2 is replaced by C/5 profile in Scenarios 6 and 13, whereas P is replaced by DST profile in Scenario 22. Behind these changes, it seems that there is a search for generalization since C/5 profile goes to lower current rates compared with C/2, and DST can be seen as a more dynamic pulse test than P. 
Even if these arguments can be used to explain some of the counter-intuitive results obtained, it should be kept in mind that this is not the full story since they overlook the situations where the expected result was actually gotten. 

Notice that based on the arguments for the counter-intuitive situations, one would expect that all the validations would present their minimum RMSE when more dynamic data (e.g. DST or P) is added to the dataset, or that some data that might interfere negatively in the parameter optimal identification (like long sequences of high current data or large datasets) is excluded, or that data tends to generalize to extreme values (like C/5 and 1C for current magnitudes, or DST for dynamic content). 
However, this does not occur consistently as seen by the many Scenarios where the estimation and validation datasets coincided in their minimum RMSE. 
This study just highlights the complex interplay of the dynamic features and richness of the estimation profiles, the nonlinear dynamics of the model, parameter generalizability and potential overfitting/underfitting during estimation.

\begin{table}[!ht]
\centering
\footnotesize
\caption{Changes in the dataset in order to achieve a minimum values of error during validation.} 
\label{Table5}
\makebox[\textwidth][c]{%
  \begin{tabular}{@{}c c c c c c c@{}}
    \hline
    Scenario & Ideal data & Validation (min)$^\dagger$ & Added & Excluded & Swapped & Validation (max)$^\dagger$ \\
    \hline
    1   & \{C/5,C/2,1C,P,DST\}  & \{/C/5,C/2,1C,P,DST\}  & -           & -         & -                  & \{C/2,P\} \\ \hline
    2   & \{C/5,C/2,1C,P\}      & \{C/5,C/2,1C,P,DST\}  & \{DST\}     & -         & -                  & \{1C,P,DST\} \\
    3   & \{C/5,C/2,1C,DST\}    & \{C/5,C/2,DST\}       & -           & \{1C\}    & -                  & \{1C,P,DST\} \\
    4   & \{C/5,C/2,P,DST\}     & \{C/5,C/2,P,DST\}     & -           & -         & -                  & \{1C,P,DST\} \\
    5   & \{C/5,1C,P,DST\}     & \{C/5,1C,P,DST\}     & -           & -         & -                  & \{C/2,P\} \\
    6   & \{C/2,1C,P,DST\}     & \{C/5,1C,P,DST\}     & -           & -         & C/2 $\rightarrow$ C/5  & \{C/2,P\} \\ \hline
    7   & \{C/5,C/2,1C\}        & \{C/5,C/2,1C\}        & -           & -         & -                  & \{1C,P,DST\} \\
    8   & \{C/5,C/2,P\}         & \{C/5,C/2,P\}         & -           & -         & -                  & \{1C,P,DST\} \\
    9   & \{C/5,C/2,DST\}       & \{C/5,C/2,DST\}       & -           & -         & -                  & \{1C,P,DST\} \\
    10  & \{C/5,1C,P\}         & \{C/5,1C,P,DST\}      & \{DST\}     & -         & -                  & \{C/5,C/2,P,DST\} \\
    11  & \{C/5,1C,DST\}       & \{C/5,1C,DST\}       & -           & -         & -                  & \{1C,P,DST\} \\
    12  & \{C/5,P,DST\}        & \{C/5,P,DST\}        & -           & -         & -                  & \{1C,P,DST\} \\
    13  & \{C/2,1C,P\}         & \{C/5,1C,P,DST\}      & \{DST\}     & -         & C/2 $\rightarrow$ C/5  & \{C/2,P\} \\
    14  & \{C/2,1C,DST\}       & \{C/2,1C,DST\}       & -           & -         & -                  & \{1C,P,DST\} \\
    15  & \{C/2,P,DST\}        & \{C/2,DST\}         & -           & \{P\}     & -                  & \{1C,P,DST\} \\
    16  & \{1C,P,DST\}        & \{1C,P,DST\}        & -           & -         & -                  & \{C/5,C/2,P,DST\} \\ \hline
    17  & \{C/5,C/2\}           & \{C/5,C/2,1C\}        & \{1C\}      & -         & -                  & \{1C,P,DST\} \\
    18  & \{C/5,1C\}           & \{C/5,C/2,1C\}        & \{C/2\}     & -         & -                  & \{1C,P,DST\} \\
    19  & \{C/5,P\}            & \{C/5,P,DST\}        & \{DST\}    & -         & -                  & \{1C,P,DST\} \\
    20  & \{C/5,DST\}          & \{C/5,P,DST\}        & \{P\}     & -         & -                  & \{1C,P,DST\} \\
    21  & \{C/2,1C\}           & \{C/2,1C,DST\}       & \{DST\}   & -         & -                  & \{1C,DST\} \\
    22  & \{C/2,P\}            & \{C/2,DST\}         & -          & -         & P $\rightarrow$ DST  & \{1C,P,DST\} \\
    23  & \{C/2,DST\}          & \{C/2,DST\}         & -          & -         & -                  & \{1C,P,DST\} \\
    24  & \{1C,P\}            & \{1C,P,DST\}        & \{DST\}   & -         & -                  & \{C/5,C/2,P,DST\} \\
    25  & \{1C,DST\}          & \{1C,P,DST\}        & \{P\}     & -         & -                  & \{C/2,P,DST\} \\
    26  & \{P,DST\}           & \{P,DST\}          & -          & -         & -                  & \{1C,P,DST\} \\ \hline
    27  & \{C/5\}              & \{C/5,C/2,1C\}        & \{C/2,1C\}  & -         & -                  & \{1C,P,DST\} \\
    28  & \{C/2\}              & \{C/2,DST\}         & \{DST\}   & -         & -                  & \{1C,P,DST\} \\
    29  & \{1C\}              & \{1C\}             & -          & -         & -                  & \{C/5\} \\
    30  & \{P\}               & \{P,DST\}          & \{DST\}   & -         & -                  & \{1C,P,DST\} \\
    31  & \{DST\}             & \{C/5,C/2,1C,P,DST\}  & \{C/5,C/2,1C,P\} & -      & -                  & \{1C,P,DST\} \\
    \hline
  \end{tabular}%
}
\\
$^\dagger$This validation (min/max) refer to the estimation datasets that minimize/maximize the validation.
\end{table}

\subsection{L4: Second Smallest Value Obtained for Each Column}

If the search is expanded from the minimum to small values in each column (condition L4), it is possible to see that the inclusion of the second smallest value per column adds 5 Scenarios to the previously mentioned 13 ones, namely \{2,10,19,21,24\}, for a total of 18 situations with relatively small errors. 
Using the same reasoning, we can extend to the next smallest value like the third, fourth and fifth ones, incorporating Scenarios \{6,17,22,25\}, \{3,13,20,31\} and \{27\}, respectively, in order to cover 27 Scenarios. These Scenarios still represent errors that are below the 10\% mark. The remaining Scenarios are \{15,18,28,30\} corresponding to error marks above 10\% that are outside the first five smallest values, for the datasets \{C/2,P,DST\}, \{C/5,1C\}, \{C/2\}, \{P\}. 
This is summarized in Table \ref{Table6}. 
In this situation, we add a set of the smallest values to have a good correspondence between estimation and validation and therefore between Cases and Scenarios (diagonal of matrix in Figure~\ref{figure6}). 
This situation portrays the occurrence of values above the minimum in the diagonal elements of the matrix while the minima appear in off-diagonal elements associated to the estimation of model parameters with a different profile than the one associated to the diagonal. 
This event can be explained in terms of the nonlinearity of the error function and possible multiple local minima of the parameter space.

\begin{table}[!ht]
\centering
\footnotesize
\caption{Other small values in the diagonal of the RMSE matrix in Figure~\ref{figure6} that are larger than their minimum column-wise.} 
\label{Table6}
\makebox[\textwidth][c]{%
  \begin{tabular}{@{}c | c | c c c@{}}
    \hline
    Smallest  & Scenarios           & Scenarios          & Corresponding       & Associated\\ 
    values    & (in diagonal)       & (off-diagonal)     & Case                & error [\%]\\
    \hline
    \multirow{2}{*}{Minimum}				
              & \{1,4,5,7,8,9,11,12,  & \multirow{2}{*}{\{15,22,23,26,29\}} 
              & \multirow{2}{*}{\{23,23,23,26,29\}}  
              & \multirow{2}{*}{\{0,0,0,0,0\}} \\
              & 14,16,23,26,29\}    &                   &                    & \\
    2nd       & \{2,10,19,21,24\}   & \{1,14,21,25\}     & \{21,21,21,29\}     & \{8,5,4,0.3\} \\
    3rd       & \{6,17,22,25\}      & \{2,4,24\}         & \{21,23,29\}        & \{11,3,2\}\\
    \multirow{2}{*}{4th} 
              & \multirow{2}{*}{\{3,13,31\}}  
              & \{5,7,8,9,10,  & \{21,17,22,26,21,   & \{7,6,6,4,8,\\
              &                   & 12,16,19,20,31\}   & 26,29,26,27,31\}    & 12,4,14,3,6\}\\
    \multirow{2}{*}{5th} 
              & \multirow{2}{*}{\{20,27\}}
              & \{3,6,11,13,17,  & \{20,21,12,21,26,   & \{6,11,16,14,6,\\
              &                   & 18,27\}            & 19,27\}             & 7,4\}\\
    Other     & \{15,18,28,30\}    & \{28,30\}          & \{28,30\}           & \{15,14\}\\
    \hline
  \end{tabular}%
}
\end{table}

It is important to notice that the minima is not the only important part but also how far the next smallest values are. Indeed, if the next small error value is relatively close to the minimum one, it might still be a reasonable error mark but the parameter error and total time requirements might be more important for this sub-optimal solution compared to the optimal one. 
For Scenario 1 (column 1 in Figure \ref{figure6}), for instance, the next error mark to the minimum appears for Case 21 with an 8\% of error difference. However, while Case 1 (the minimum) requires the use of all the available datasets, i.e. \{C/5,C/2,1C,P,DST\}, Case 21 resorts to dataset \{C/2,1C\} at the price of 8\% more of output error. 
This is exemplified in the last 3 columns of Table~\ref{Table6} where the minimal dataset for Scenario 2 corresponds to the 2nd minimum (row 2, column 2), for which the corresponding Case is 21 (row 2, column 3) with associated 8\% error (row 2, column 4). 
A similar reasoning can be used for the other Scenarios. 
In most situations, the required data will be reduced by 1 to 3 datasets since all situations above Case 16 (see Figure~\ref{figure6}) require the full 5 datasets (Case 1), 4 datasets (Cases 2 to 6) or 3 datasets (Cases 7 to 16) and data reduction can go up to 2 datasets (Cases 17 to 26) or 1 dataset (Cases 27 to 31) if an error within the 5 smallest values is acceptable (below 15\% of error). An exception is Scenario 11, where the minimum error is found for Case 11 and the shortest dataset that can be obtained within the first 5 smallest errors is for Case 12, with the 5th smallest error of 16\% and still keeping 3 datasets (from \{C/5,1C,DST\} to \{C/5,P,DST\}). 
In general, dataset reduction is dependent on the original dataset length, where longer datasets are more prone to be reduced with small errors than shorter datasets. 
In the limit where only 1 dataset is used (Cases 27 to 31), no further reduction is possible and in some Cases, the corresponding error is above the 5 smallest errors for that column (for Scenarios 28 and 30 for instance). 
These notions are naturally exploited below when a cost function is used to find the best combination of conditions to optimize a given set of metrics.

\subsection{L5: Maximum Value Obtained for Each Column}

In terms of maximum error (condition L5), the situation that incurs in the largest error most of the times (21 out of 31) corresponds to the estimation of parameters with the dataset \{1C,P,DST\} (see red entries with values of 1 in row 16 in Figure~\ref{figure6} and last column of Table \ref{Table5}). 
However, when validated with the same data of \{1C,P,DST\}, the minimum is achieved compared to the use of other parameter estimates from other profiles (see green entry with value of 0 in column 16 in Figure~\ref{figure6}). 
This profile is then more susceptible to errors during validation with other profiles, but it will reach a minimum compared to other possible parameter estimates. 
The other maximum error values can be counted as 4 times when using the \{C/2,P\} profile, followed by 3 times when using the \{C/5,C/2,P,DST\} profile. 
The datasets with a single occurrence each are \{1C,DST\}, \{C/2,P,DST\} and \{C/5\}.

\subsection{L6: Parameter Errors}

Table~\ref{Table7} shows the component-wise relative error of each parameter component in the parameter vector Eq.~\eqref{eq:12} as defined in Eq.~\eqref{eq:16} shown in column 2 to 10, where $\delta_{\theta_j}$ (with $j = 1,\ldots,9$) in Table~\ref{Table7} just specifies the considered parameter component, the average parameter error in column 11, 
the Euclidean distance of the parameter vector errors (as in Eq.~\eqref{delta}) in column 12 $\delta_{\rm dist}$. The  important notion is the distance between the identified parameters and their nominal values. 
Starting from the component-wise errors, we see that minimum errors appear for the parameters \{$\alpha^-, \alpha^+, Q^-, Q^+, d^-, d^+, {\rm SOC}_0^-, {\rm SOC}_0^+, R_0$\} in the Cases \{19, 28, 28, 25, 26, 4, 3, 19, 21\}, respectively, with errors of \{1.44, 0.47, 0.08, 0.24, 2.94, 0.74, 0.70, 0.07, 0.88\}\%, respectively. 
Thus, the maximum error in a parameter component is around 3\% for $d^-$. 
Most component minimum errors show up for datasets with 1 or 2 profiles (Cases 17 to 31) except for the errors for $d^+$ and ${\rm SOC}_0^-$ for Cases 4 and 3, respectively. 
The minimum of the average of these errors for each Case (column 11 in Table~\ref{Table7}) corresponds to 8.45\% for Case 4. 
Even if this gives an idea of the level of error of each parameter component, we also look at the Euclidean distance between the estimated parameter vector and the nominal one (column 12 in the table). 

In this Case, the minimum distance is also found in Case 4 with $\delta_{\rm dist} = 562.56$, i.e. dataset \{C/5,C/2,P,DST\}. On the other hand, the maximum error occurs for Case 6 with $\delta_{\rm dist} = 2619.77$, i.e. dataset \{C/2,1C,P,DST\}. Therefore, the convergence of the parameter vector is related to the usage of more dynamic data including P and DST.

Regarding the estimation errors of battery parameters, our parameter errors results indicate an interesting phenomenon. For instance, in Case 1, the model exhibited the smallest voltage error; however, its overall parameter error was greater than that of Case 4. This finding implies that achieving the minimum voltage error does not necessarily correspond to minimizing the battery parameter error. The underlying issue is that the parameters in the electrochemical model are prone to overfitting. Even though we grouped the battery parameters, there may still exist unmodeled, latent correlations among them. Consequently, the model can produce a very low overall voltage output error while still incurring high parameter errors. In other words, different parameter sets may lead to similar model outputs, which poses a significant challenge for parameter estimation algorithms in identifying the correct optimization direction. This observation underscores the inherent complexity of electrochemical models and the challenges associated with accurate parameter estimation.

\begin{table}[!ht]
\caption{Relevant metrics for the error between estimated and nominal parameter vectors both component-wise and Euclidean distance.}

\label{Table7}
\hspace{0cm}
{\small
\begin{tabular}{c | c c c c c c c c c c c}
\hline

\multirow{ 2}{*}{Cases}   

        &\multirow{ 2}{*}{$\delta_{\alpha^-}$}
        &\multirow{ 2}{*}{$\delta_{\alpha^+}$}
        &\multirow{ 2}{*}{$\delta_{Q^-}$}	
        &\multirow{ 2}{*}{$\delta_{Q^+}$}	
        &\multirow{ 2}{*}{$\delta_{d^-}$}	
        &\multirow{ 2}{*}{$\delta_{d^+}$}
        &\multirow{ 2}{*}{$\delta_{{\rm SOC}_0^-}$}
	&\multirow{ 2}{*}{$\delta_{{\rm SOC}_0^+}$}
	&\multirow{ 2}{*}{$\delta_{R_0}$}
	&\multirow{ 2}{*}{$\bar{\delta}_{\theta}$}
	&\multirow{ 2}{*}{$\delta_{\rm dist}$} \\
&&&&&&&&&&&\\
\hline
1	&56.19	&19.08	&1.80 	&2.14	&33.24 	&31.53	&1.14 	&33.75 	&25.28    &22.68 	&1807.17\\ \hline
2	&58.93	&32.57	&1.74	&0.77	&64.50	&53.64	&2.52	&41.38	&64.34    &35.60	&1939.08\\
3	&17.77	&8.62	&0.79	&1.90	&74.42	&44.91	&0.70	&63.26	&66.28    &30.96	&618.00\\
4	&11.47	&3.22	&3.51	&1.87	&8.81	&0.74	&4.07	&30.66	&11.65    &8.45	    &562.56\\
5	&59.77	&22.97	&0.11	&0.83	&24.49	&22.40	&1.62	&20.03	&18.42    &18.96	&1907.31\\
6	&70.22	&60.40	&8.08	&2.58	&39.48	&14.47	&5.56	&22.07	&44.41    &29.70	&2619.77\\  \hline
7	&24.32	&35.85	&5.79	&1.39	&3.50	&6.99	&5.27	&5.63	&19.40    &12.02	&1196.17\\
8	&4.57	&32.78	&6.76	&0.73	&25.41	&30.20	&6.88	&16.04	&55.35    &19.86	&964.17\\
9	&16.46	&18.23	&4.17	&3.27	&16.66	&29.49	&3.94	&67.36	&26.96    &20.73	&843.40\\
10	&63.38	&12.66	&4.61	&0.87	&11.87	&21.62	&2.82	&26.74	&4.29     &16.54	&2045.64\\
11	&25.09	&52.13	&1.50	&0.52	&272.75	&22.53	&1.79	&4.16	&95.48    &52.89	&1258.05\\
12	&19.36	&11.56	&5.88	&0.30	&30.72	&19.76	&5.77	&8.56	&24.13    &14.00	&899.95\\
13	&55.93	&9.77	&0.40	&2.94	&33.72	&10.87	&1.79	&86.94	&25.91    &25.36	&1777.22\\
14	&26.55	&16.58	&2.78	&2.90	&75.74	&35.00	&3.59	&83.68	&39.23    &31.78	&984.33\\
15	&15.10	&31.57	&10.33	&2.30	&14.57	&16.29	&9.86	&21.57	&21.60    &15.91	&1367.05\\
16	&59.23	&38.55	&10.93	&0.68	&27.80	&35.15	&6.53	&11.81	&32.18    &24.76	&2300.05\\  \hline
17	&33.59	&32.73	&5.80	&1.82	&164.10	&35.53	&5.19	&18.57	&116.8    &46.01	&1375.43\\
18	&32.83	&59.68	&1.66	&5.07	&18.21	&5.29	&1.28	&152.45	&30.19    &34.07	&1621.52\\
19	&1.44	&35.91	&7.67	&2.61	&25.78	&53.31	&7.37	&0.07	&69.34    &22.61	&1103.08\\
20	&40.55	&29.59	&9.11	&1.70	&3.90	&26.14	&8.03	&53.61	&2.27     &19.43	&1699.46\\
21	&41.42	&17.47	&1.08	&2.55	&8.13	&4.46	&1.97	&90.49	&0.88     &18.72	&1362.04\\
22	&45.60	&46.36	&5.60	&2.13	&5.76	&27.01	&6.39	&8.42	&39.65    &20.77	&1781.27\\
23	&33.64	&17.94	&2.89	&0.95	&9.17	&64.72	&3.73	&15.12	&1.22     &16.60	&1145.34\\
24	&32.73	&13.39	&8.32	&0.51	&13.87	&28.89	&5.38	&15.71	&0.88     &13.30	&1379.13\\
25	&39.03	&20.77	&1.80	&0.24	&106.40	&38.02	&1.04	&40.51	&41.49    &32.14	&1287.57\\
26	&34.52	&15.23	&5.82	&1.08	&2.94	&10.64	&5.26	&2.99	&26.08    &11.62	&1279.52\\ \hline
27	&60.61	&50.27	&3.39	&2.69	&12.55	&51.30	&2.66	&5.69	&30.08    &24.36	&2155.08\\
28	&71.68	&0.47	&0.08	&4.52	&674.07	&27.20	&1.62	&116.70	&128.82   &113.91	&2281.95\\
29	&39.41	&15.12	&1.17	&0.43	&502.06	&32.31	&3.41	&25.23	&104.77   &80.43	&1263.15\\
30	&16.67	&54.18	&11.64	&5.15	&11.04	&18.31	&10.48	&81.73	&43.14    &28.04	&1785.37\\
31	&55.41	&13.79	&3.60	&1.99	&18.92	&11.21	&2.79	&5.84	&6.26     &13.31	&1796.00\\
\hline 
\end{tabular}

}

\end{table}

\subsection{L7: Overall Best Datasets for Given Performance Metrics}

After analyzing the results associated to estimating and validating the obtained parameter vectors with a given set of data, we are interested in finding the optimal parameters that are able to minimize: a) the RMSE across data profiles, b) the distance between the identified and nominal parameter values, c) the time required for parameterization (experimental and optimization). Now that we know extreme cases of minimum and maximum for the considered relevant metrics, we want to find optimal solutions to the minimum dataset problem that are appropriate weighted combinations of these different metrics in accordance to what a user may prioritize (condition L7). 
In order to do so, we craft the following cost function
\begin{equation}
J = \alpha \bar{e}_y + \beta \bar{e}_\theta + \gamma \bar{T}_{\rm total}
\end{equation}
where the bar notation $\bar{\chi}$ implies the normalization of the given variable with respect to the its minimum $\chi_{\rm min}$ and maximum $\chi_{\rm max}$ values, i.e.
\begin{equation}
\bar{\chi} = \frac{\chi - \chi_{\rm min}}{\chi_{\rm max} - \chi_{\rm min}},
\end{equation}
$e_y$ is the  output $\rm RMSE$ for Scenario 1, which encompasses all five fundamental operating conditions, 
$e_\theta$ is the Euclidean distance between the estimated optimal parameter $\theta^*$ and its nominal value $\theta_{\rm true}$, 
$T_{\rm total}$ is the total time, 
and cost function parameters $\alpha, \beta, \gamma \in [0,1]$ (with $\alpha + \beta + \gamma = 1$) are used to test different options (from O1 to O7) to emphasize a given set of conditions, namely to minimize:
\begin{enumerate}
\item[O1] output error with $\alpha = 1$, $\beta = \gamma = 0$ (output error focus);  
\item[O2] parameter vector error with $\alpha = \gamma = 0$, $\beta = 1$ (parameter error focus) ;
\item[O3] total time with $\alpha = \beta = 0$, $\gamma = 1$ (time focus); 
\item[O4] combination of [O1] and [O2] with $\alpha = \beta = 1/2$, $\gamma = 0$ (output and parameter error focus); 
\item[O5] combination of [O1] and [O3] with $\alpha = \gamma = 1/2$, $\beta = 0$ (output error-time focus); 
\item[O6] combination of [O2] and [O3] with $\beta = \gamma = 1/2$, $\alpha = 0$ (parameter error-time focus); 
\item[O7] combination of [O1], [O2] and [O3] with $\alpha = \beta = \gamma = 1/3$ (balanced approach).
\end{enumerate}

Here, we either pick extreme conditions or uniform weights but different cost function parameters could potentially be used in the case of different user requirements. 
The obtained results are shown in Table~\ref{Table8}.

\begin{table}[!ht]
\centering
\footnotesize
\caption{Relevant metrics for this study, \(e_y\), \(e_\theta\), \(T_{\mathrm{total}}\) and O1--O7 results.}
\label{Table8}
\begin{tabular}{c|ccc|ccccccc}
\toprule
\multirow{2}{*}{\textbf{Case}} 
 & \multicolumn{3}{c|}{\textbf{Original Metrics}} 
 & \multicolumn{7}{c}{\textbf{Cost Function Results}}\\
\cline{2-11}
& $e_y$ [V]& $e_\theta$[-] & $T_{\rm total}[h]$ 
& O1 & O2 & O3 & O4 & O5 & O6 & O7\\
\midrule
1  & 0.0259 & 1807.17 & 55.0 & 0.000 & 0.605 & 1.000 & 0.303 & 0.500 & 0.803 & 0.535 \\
2  & 0.0280 & 1939.08 & 50.6 & 0.107 & 0.669 & 0.915 & 0.388 & 0.511 & 0.792 & 0.564 \\
3  & 0.0285 & 618.00  & 23.2 & 0.132 & 0.027 & 0.388 & 0.080 & 0.260 & 0.208 & 0.182 \\
4  & 0.0406 & 562.56  & 52.0 & 0.746 & 0.000 & 0.942 & 0.373 & 0.844 & 0.471 & 0.563 \\
5  & 0.0293 & 1907.31 & 50.2 & 0.173 & 0.654 & 0.908 & 0.414 & 0.541 & 0.781 & 0.578 \\
6  & 0.0323 & 2619.77 & 44.0 & 0.325 & 1.000 & 0.788 & 0.663 & 0.557 & 0.894 & 0.704 \\
7  & 0.0294 & 1196.17 & 18.8 & 0.178 & 0.308 & 0.304 & 0.243 & 0.241 & 0.306 & 0.263 \\
8  & 0.0377 & 964.17  & 51.0 & 0.599 & 0.195 & 0.923 & 0.397 & 0.761 & 0.559 & 0.572 \\
9  & 0.0283 & 843.40  & 20.2 & 0.122 & 0.136 & 0.331 & 0.129 & 0.227 & 0.234 & 0.196 \\
10 & 0.0299 & 2045.64 & 45.8 & 0.203 & 0.721 & 0.823 & 0.462 & 0.513 & 0.772 & 0.582 \\
11 & 0.0287 & 1258.05 & 18.4 & 0.142 & 0.338 & 0.296 & 0.240 & 0.219 & 0.317 & 0.259 \\
12 & 0.0343 & 899.95  & 47.5 & 0.426 & 0.164 & 0.856 & 0.295 & 0.641 & 0.510 & 0.482 \\
13 & 0.0298 & 1777.22 & 39.6 & 0.198 & 0.590 & 0.704 & 0.394 & 0.451 & 0.647 & 0.497 \\
14 & 0.0304 & 984.33  & 12.1 & 0.228 & 0.205 & 0.175 & 0.217 & 0.202 & 0.190 & 0.203 \\
15 & 0.0375 & 1367.05 & 41.0 & 0.589 & 0.391 & 0.731 & 0.490 & 0.660 & 0.561 & 0.570 \\
16 & 0.0449 & 2300.05 & 39.1 & 0.964 & 0.845 & 0.694 & 0.905 & 0.829 & 0.769 & 0.834 \\
17 & 0.0297 & 1375.43 & 15.9 & 0.193 & 0.395 & 0.248 & 0.294 & 0.221 & 0.322 & 0.279 \\
18 & 0.0284 & 1621.52 & 14.0 & 0.127 & 0.515 & 0.212 & 0.321 & 0.170 & 0.364 & 0.285 \\
19 & 0.0353 & 1103.08 & 43.0 & 0.477 & 0.263 & 0.769 & 0.370 & 0.623 & 0.516 & 0.503 \\
20 & 0.0287 & 1699.46 & 15.6 & 0.142 & 0.553 & 0.242 & 0.348 & 0.192 & 0.398 & 0.312 \\
21 & 0.0275 & 1362.04 & 7.8  & 0.081 & 0.389 & 0.092 & 0.235 & 0.087 & 0.241 & 0.187 \\
22 & 0.0456 & 1781.27 & 36.9 & 1.000 & 0.592 & 0.652 & 0.796 & 0.826 & 0.622 & 0.748 \\
23 & 0.0424 & 1145.34 & 9.2  & 0.837 & 0.283 & 0.119 & 0.560 & 0.478 & 0.201 & 0.413 \\
24 & 0.0399 & 1379.13 & 34.8 & 0.710 & 0.397 & 0.612 & 0.554 & 0.661 & 0.505 & 0.573 \\
25 & 0.0424 & 1287.57 & 7.3  & 0.837 & 0.352 & 0.083 & 0.595 & 0.460 & 0.218 & 0.424 \\
26 & 0.0315 & 1279.52 & 36.3 & 0.284 & 0.349 & 0.640 & 0.317 & 0.462 & 0.495 & 0.424 \\
27 & 0.0325 & 2155.08 & 11.1 & 0.335 & 0.774 & 0.156 & 0.555 & 0.246 & 0.465 & 0.422 \\
28 & 0.0327 & 2281.95 & 4.8  & 0.345 & 0.836 & 0.035 & 0.590 & 0.190 & 0.436 & 0.405 \\
29 & 0.0403 & 1263.15 & 3.0  & 0.731 & 0.341 & 0.000 & 0.536 & 0.366 & 0.171 & 0.357 \\
30 & 0.0323 & 1785.37 & 31.8 & 0.325 & 0.594 & 0.554 & 0.459 & 0.440 & 0.574 & 0.491 \\
31 & 0.0300 & 1796.00 & 4.4  & 0.208 & 0.600 & 0.027 & 0.404 & 0.118 & 0.314 & 0.278 \\
\bottomrule
\end{tabular}
\end{table}

The optimal operating condition for each conditions is summarized in Table~\ref{Table9}. From the results in Table~\ref{Table9}, it can be observed that if we only consider minimizing the output voltage error (O1), Case 1 \{C/5,C/2,1C,P,DST\} is the optimal choice as it encompasses all operating conditions. However, it also has the longest computation time (55 hours). Interestingly, although the model voltage output error for Case 1 is the smallest, its parameter error is not the lowest. This highlights the significant challenges in electrochemical model parameter estimation and the issue of over-parameterization, where multiple combinations of parameters can result in similarly accurate model outputs. In other words, large parameter errors may still yield small model voltage output errors. This discrepancy may also be attributed to differences in parameter sensitivity, as some parameters exhibit inherently low sensitivity and are therefore difficult to estimate accurately. In addition, the feasible range of battery parameters can further influence the parameter estimation results.

If we focus solely on minimizing parameter error (O2), Case 4 \{C/5,C/2,P,DST\} is the optimal choice, but it has relatively large output voltage errors and a long computation time (52 hours). When only considering the shortest time (O3), Case 29 \{1C\} emerges as the best option since it includes only the 1C operating condition, resulting in the shortest runtime (3 hours). However, its model voltage output error is relatively high (0.0403 V).

When both model voltage output error and parameter error are considered, Case 3 \{C/5,C/2,1C,DST\} is the optimal choice. If the focus is on minimizing model output error and time (O5), Case 21 \{C/2,1C\} is the best option, aligning with the results in Section 3.1. Conversely, if the priority is to minimize parameter error and computation time, Case 29 \{1C\}	 becomes the optimal choice due to its shortest runtime (3 hours). When comprehensively considering model voltage output error, parameter error, and computation time, Case 3 \{C/5,C/2,1C,DST\} is the best choice.

Based on the above results, the optimal operating condition can be selected according to specific needs. For example, if the primary concern is model voltage output error and computation time (O5), Case 21 \{C/2,1C\}  is the recommended option, as this aligns with the most common practical requirements. In many real-world scenarios, the exact range of parameter values, potential over-parameterization, and parameter sensitivity are of secondary importance; the primary focus is instead on reducing the model’s voltage error. However, if more accurate battery parameter information is needed, such as for explaining internal resistance increases caused by aging, Case 4\{C/5,C/2,P,DST\} should be considered despite the increased computation time. Researchers can select the most suitable operating condition based on their specific requirements.

In fact, the findings of this study can provide direct guidance for such applications. For instance, in model-based SOC estimation methods, the SOC accuracy is strongly influenced by the model’s voltage prediction error. Therefore, the optimal current combinations identified for minimizing voltage error (e.g., O1 and O5 in Table~\ref{Table9}) are particularly well-suited for SOC estimation tasks. Conversely, in the context of SOH re-calibration, capacity-related parameters play a more critical role. Hence, current combinations that yield lower errors in capacity-related parameters (e.g., O2 and O6 in Table~\ref{Table9}) are preferable for SOH estimation. Accordingly, our systematic evaluation of 31 operating condition combinations offers practical insights for selecting appropriate test profiles tailored to different application goals (SOC vs. SOH).

\begin{table}[!ht]
\centering
\footnotesize
\caption{Optimal datasets for different minimization metrics.} 
\label{Table9}
\makebox[\textwidth][c]{%
  \begin{tabular}{@{}l | c c l c c c@{}}
    \toprule
    \textbf{Conditions}  & \textbf{Minimum $J$} & \textbf{Case} & \textbf{Suggested data} & \textbf{$e_y$ [V]} & \textbf{$e_\theta$ [-]} & \textbf{$T_{\rm total}$ [h]} \\
    \midrule
    O1: Output error             & 0.000 & 1  & \{C/5,C/2,1C,P,DST\} & 0.0259 & 1807.17 & 55.0 \\
    O2: Parameter error          & 0.000 & 4  & \{C/5,C/2,P,DST\}    & 0.0406 & 562.56  & 52.0 \\
    O3: Time requirement         & 0.000 & 29 & \{1C\}              & 0.0403 & 1263.15 & 3.0  \\
    O4: Output and parameter error & 0.080 & 3  & \{C/5,C/2,1C,DST\}  & 0.0285 & 618.00  & 23.2 \\
    O5: Output error-time        & 0.087 & 21 & \{C/2,1C\}         & 0.0275 & 1362.04 & 7.8  \\
    O6: Parameter error-time     & 0.171 & 29 & \{1C\}              & 0.0403 & 1263.15 & 3.0  \\
    O7: Balanced                 & 0.182 & 3  & \{C/5,C/2,1C,DST\}   & 0.0285 & 618.00  & 23.2 \\
    \bottomrule
  \end{tabular}%
}
\end{table}

\subsection{Limitation}

This study focuses exclusively on lithium-ion batteries utilizing NMC materials, and the conclusions drawn are specifically applicable to these systems. Although the comprehensive analytical methodology presented here has been successfully demonstrated for NMC-based batteries, it is equally applicable to other battery chemistries, including Lithium Iron Phosphate (LFP) and other batteries. By applying the same approach to LFP and other type batteries, researchers may streamline testing protocols and obtain corresponding performance insights. The number of PSO iterations may also have a potential impact on the results. In this study, we adopted a commonly used setting of 300 iterations for the analysis. Using a larger number of iterations could reduce the error for certain operating conditions, but it would also lead to a longer computational time.

\section{Conclusions}

The parameter estimation of electrochemical models has long been a significant challenge in their practical application. This study investigates the influence of different current conditions on battery model parameter estimation for NMC lithium-ion batteries , aiming to identify the optimal current conditions for parameter estimation. Based on five basic current conditions (C/5, C/2, 1C, Pulse, DST), 31 combinations of current conditions were generated. These 31 combinations were used both for parameter estimation and for validating the estimation results, resulting in a total of 961 evaluation outcomes. The analysis focuses on three dimensions: model voltage output error, battery parameter error, and computation time. The results indicate the following. To achieve the minimum model voltage output error, all five basic current conditions ({C/5, C/2, 1C, Pulse, DST}) should be used as input for parameter estimation.
For minimizing parameter estimation error, the optimal set of current conditions is {C/5, C/2, Pulse, DST}.
The shortest computation time is achieved with the current condition {1C}.
When considering both output error and parameter error, the optimal set of current conditions is {C/5, C/2, 1C, DST}.
For minimizing model voltage output error and computation time simultaneously, the best current condition set is {C/2, 1C}.
When both parameter error and computation time are prioritized, {1C} is the optimal choice.
The comprehensive optimal set of current conditions, considering model voltage output error, parameter error, and computation time, is {C/5, C/2, 1C, DST}.
Researchers can select the most appropriate current conditions based on their specific requirements. Future studies could incorporate sensitivity analysis of battery model parameters to identify the most sensitive current conditions corresponding to each parameter.

\section*{Acknowledgments}
This work was supported by the Research Foundation - Flanders (FWO) (grant numbers 1252326N, 2025).

\section*{Author contribution} 
Feng Guo: Conceptualization, Methodology, Software, Formal analysis, Investigation, Visualization, Writing – original draft.
Luis D. Couto: Conceptualization,  Methodology, Visualization, Writing – original draft, Writing – review \& editing. Khiem Trad: Conceptualization, Writing – review \& editing. Grietus Mulder: Writing – review \& editing.
Keivan Haghverdi: Writing – review \& editing.
Guillaume Thenaisie: Writing – review \& editing.

\section*{Data Availability}
The implementation of the models used in this paper can be accessed at 
\url{https://github.com/FrankSuperG/CPG-SPMT}. 

\section*{Conflict of Interest}
The authors declare that they have no conflict of interest.

 \bibliographystyle{elsarticle-num} 
 \bibliography{cas-refs}
\end{document}